\documentclass[aps,prd,preprint,superscriptaddress,nofootinbib,showkeys,showpacs]{revtex4}

\usepackage{epsfig}
\usepackage{amsmath}
\usepackage{dcolumn}
\usepackage{multirow}

\newcommand{\Uppsala}{Division of High Energy Physics, Uppsala University. S-75121 Uppsala, Sweden}

\begin{document}


\title{Including Systematic Uncertainties in 
Confidence Interval Construction for Poisson Statistics}



\author{J. Conrad}
\author{O. Botner}
\author{A. Hallgren}
\author{C. P\'erez~de~los~Heros}

\affiliation{\Uppsala}

\date{\today}

\begin{abstract}
One way to incorporate systematic uncertainties into the calculation
of confidence intervals is by integrating over probability density
functions parametrizing the uncertainties. In this note we
present a development of this method which takes into account
uncertainties in the prediction of background processes, uncertainties
in the signal detection efficiency and background efficiency and
allows for a correlation between the signal and background detection
efficiencies. We implement this method with the Likelihood Ratio (usually 
denoted as Feldman \& Cousins) approach with and without conditioning.\\
We present studies of coverage for the Likelihood Ratio and Neyman
ordering schemes. In particular, we present two different types of
coverage tests for the case where systematic uncertainties are included.
To illustrate the method we show the relative effect of including systematic
uncertainties the case of dark matter search as performed by modern neutrino telescopes.

\end{abstract}



\maketitle

\section{\label{sec:Intro} Introduction}

 A limit on , or a measurement of , a physical quantity at a given confidence level is 
usually set by comparing a number of detected events, $n_o$, with the number  
of expected events from the known background sources contributing to the physical process 
in question, $n_b$. How 'compatible' these numbers are determines how much 
room there is for new processes, ie., for a signal. How well do the number of 
observed events and expected background compare, strongly depends on the 
systematic uncertainties present in the measurement. Systematic uncertainties must, therefore, 
be taken into account in the limit or confidence belt calculation that is finally published.\par

 Traditionally, confidence limits are set using a Neyman construction~\cite{Neyman:1937a}. 
This is a purely frequentist method. G.~Feldman and 
R.~Cousins~\cite{Feldman:1998a} have proposed an improved method to construct
confidence intervals based on likelihood ratios, a method allready known in statistics and originally described in \cite{Kendall:91}. Still, this method is based on the original Neyman construction, and needs to be extended to incorporate systematic uncertainties in the measurement. Along this line, a modification of the Neyman method that incorporates  systematic uncertainties in the experimental signal efficiency has been proposed  by V.~Highland \& R.~Cousins~\cite{Cousins:1992a}. These authors use a ``semi''- Bayesian approach where an average over the probability
distribution of the experimental sensitivity (and its uncertainty) is performed. By construction, the method is of limited accuracy in the limit of high relative systematic uncertainties. \par

Recently, an entirely frequentist approach has been proposed 
for uncertainty in the background rate prediction~\cite{Rolke:2001a}. 
That approach is based on a two-dimensional confidence belt construction and likelihood ratio hypothesis 
testing and treats the uncertainty in the background as a statistical uncertainty rather than as a systematic one.\par

The interest aroused recently in the High Energy Physics community about the
many open issues on setting limits and quoting confidence levels is 
stresseded by the organization of devoted workshops on the
subject. We refer the reader to the proceedings of the recent
workshops at CERN~\cite{CERN:00a}, FERMILAB~\cite{FERM:00a} and Durham~\cite{DURHAM:02} for a review of the status of the field.\par

 In this paper we extend the method of confidence belt construction proposed
in~\cite{Cousins:1992a} to include systematic uncertainties 
both in the signal and background efficiencies as well as theoretical uncertainties in the background prediction.
The proposed method allows as well to use newer ordering schemes. A recent attempt to include systematic uncertainty in the background prediction in a similar manner has been presented in \cite{Giunti:1998xv}. 
The paper is organized as follows. In section~\ref{sec:rev} we give a short review of
the confidence belt construction schemes which we will use. In section~\ref{sec:unc} we
describe how to include the systematic uncertainties, in section~\ref{sec:cos} we discuss how 
the confidence belt construction is performed and present some selected results. We compare the results of this method with other methods to include systematics in section~\ref{sec:comparison}. 

We introduce the tests 
of coverage performed in section~\ref{sec:cov} and present an example
based on data from the AMANDA neutrino experiment in section~\ref{sec:app}.

\section{\label{sec:rev} The Construction of Confidence Intervals}
The frequentist construction of confidence intervals is described in
detail elsewhere~\cite{Eadie:1982a}. Here we will give just a short review.\par
\begin{figure}[t]
\begin{minipage}[t]{0.92\linewidth}
\epsfig{file=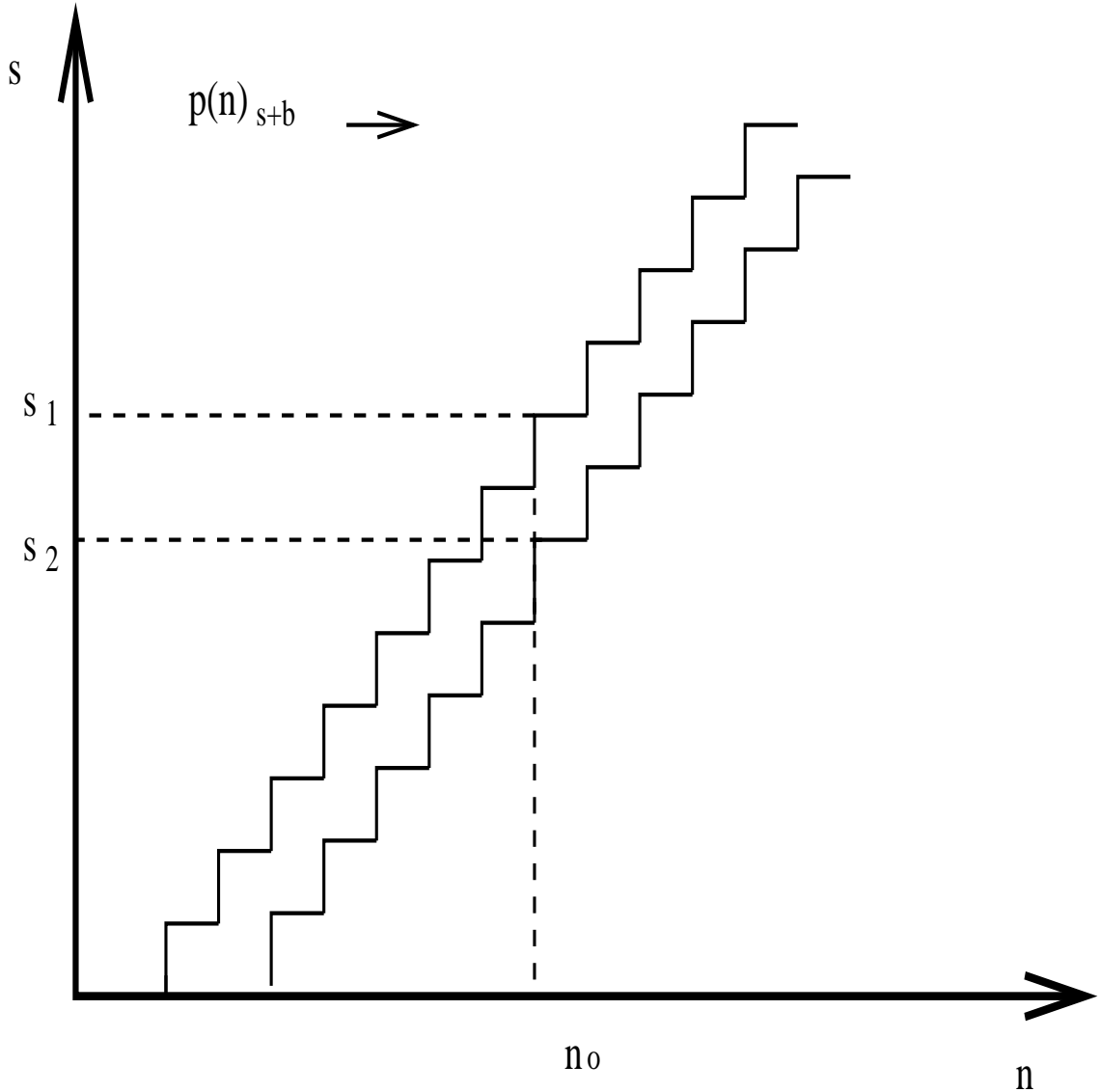,height=\linewidth,width=\linewidth}
\caption{Illustration of the confidence belt construction. On the x
axis are the possible experimental outcomes (number of events), on the
y axis the parameter of the pdf ($s$). In this case a Poisson PDF was assumed.}
\label{fig:fc}
\end{minipage}\hfill
\end{figure}

Let us consider a Poissonian probability density function (PDF), $p(n)_{s+b}$, for a fixed but unknown signal, $s$, in the presence of
a known background with mean $b$. For every value of
$s$ we can find two values $n_1$ and $n_2$ such that 
\begin{equation}
\sum_{n'=n_1}^{n_2} p(n')_{s+b} = 1 - \alpha
\label{eq:neyman0}
\end{equation}
where $1-\alpha$ denotes the confidence level (usually quoted as 
a 100(1-$\alpha$)\% confidence interval).
Since we assume a Poisson distribution, the equality will generally
not be fulfilled exactly.  A set of intervals
$[n_1(s+b,\alpha),n_2(s+b,\alpha)]$ is called a {\em confidence
belt\/}. Graphically, upon a measurement, $n_o$, the {\em confidence interval\/} $[s_1,s_2]$ is determined by the intersection of the vertical line
drawn from the measured value $n_o$ and the boundary of the confidence
belt.  This is illustrated in figure~\ref{fig:fc}.  The probability
that the confidence interval will contain the true value $s$ is $1 -
\alpha$, since this is true for all $s$ per construction. 

The choice of the $n_1$ and $n_2$ is, however, not unique to define the
confidence belt. An additional criterion has to be applied. The
choices originally proposed by Neyman~\cite{Neyman:1937a} are 
\begin{equation}
\sum_{n'=0}^{n_1} p(n')_{s+b} = \sum_{n'=n_2}^{\infty} p(n')_{s+b} = \frac{1 - \alpha}{2}
\label{eq:neyman1}
\end{equation}
for central confidence intervals, and
\begin{equation}
\sum_{n'=0}^{n_1} p(n')_{s+b} = 1 - \alpha
\label{eq:neyman2}
\end{equation}
for upper confidence limits.
This method presents certain drawbacks in the case of small samples and, in particular, 
can yield null results (in the sense that the algorithm gives no
answer) in the case  when no events have 
been observed. Also, the decision of quoting a measurement (that is, a central 
confidence interval) or an upper limit might not be straightforward before performing an experiment. 
\subsection{Likelihood ratio ordering}
To solve this problem, a modification of the  Neyman method has been proposed ~\cite{Kendall:91} ~\cite{Feldman:1998a} that is based on a more rationalized ordering scheme of the elements in the sum in equation (\ref{eq:neyman0}), based on likelihood ratios. This approach automatically provides
central confidence intervals when motivated and upper limits when
necessary, therefore it is often denoted as the ``unified approach''.
Instead of using the choices given
in the previous section, the following ordering scheme is applied in solving equation (\ref{eq:neyman0}):\\
For each $n$ the $s_{best}$ is found which maximizes the
likelihood $\mathcal{L}(n)_{s+b}$. In case of a simple Poissonian
distribution with known background, $s_{best}$ is given by $max(0,n-b)$.
Then for a fixed $s$ the ratio
\begin{equation}
R(s,n)_{\mathcal{L}} = \frac{\mathcal{L}_{s+b}(n)}{\mathcal{L}_{s_{best}+b}(n)}
\end{equation}
is computed for each $n$, and all $n$'s are consequently ranked
according to the value of this ratio. Values of $n$ are included in
the confidence belt starting with the $n$ with the 
highest rank (largest $R_{\mathcal{L}}$) and then decreasing rank 
until $\sum_{n=n_1}^{n_2} p(n)_{s+b} = 1 - \alpha$. 
After the confidence belt has been constructed in this way, the 
confidence interval $[s_1,s_2]$ is found as described in the previous
section. Note that this ordering principle is a standard method within
the theory of likelihood ratio tests ~\cite{Kendall:91}. 

\par
This approach has some undesired features as well. There is a background dependence of
the upper limit in case of less events observed than expected from 
background. 
This can lead to situations where measurements with higher background
give a better limit, a clearly undesireable effect.  
B.~Roe \& M.~Woodroofe~\cite{Roe:1999a} proposed a solution 
to this problem which we briefly describe next.

\subsection{Conditioning}
A variation of the classical method of constructing confidence belts
is to use the fact that, given an observation $n_o$, it is
known that the background can not have been larger than
$n_o$ itself. To incorporate this knowledge into the PDF, the authors
in~\cite{Roe:1999a} have proposed the following modification:
\begin{equation}
 q^{n_{o}}_{s+b}(n) = \left\{ \begin{array}{ll}
    \frac{\displaystyle p(n)_{s+b}}{\displaystyle \sum_{n'=0}^{n_o} p(n')_b}                      & \mbox{if $n \leq n_o$}\\
 \\
    \frac{\displaystyle \sum_{n'=0}^{n_o} p(n')_b
    p(n-n')_s}{\displaystyle \sum_{n'=0}^{n_o} p(n')_b}  & \mbox{if $\ n > n_o$}
   \end{array}
\right.
\end{equation}

The likelihood ratio ordering can then be applied with this new PDF. 
Note that in this case the PDF is dependent on the number of observed
events. This approach solves the background dependence of the upper
limit: a limit set when no events are observed
stays constant at a value of 2.44 independent of the expected
background (which agrees with the result of the original likelihood
ordering for no events observed and no expected background). 
However, this method does not 
satisfy all the requirements of proper coverage~\cite{Zech:2001a} and
has problems when applied to the case of a Gaussian distribution
with boundaries ~\cite{Cousins:2000a}.
An extension  based on a Bayesian approach with tests of
coverage can be found in~\cite{Roe:2001a}.

\section{\label{sec:unc} The inclusion of systematic uncertainties }
The way of incorporating systematic uncertainties into the confidence
belt construction presented in this paper does not
affect the particular ordering scheme. Instead, it takes into
account the systematic uncertainties by assuming (or if possible
determining) a PDF which parameterizes our knowledge about the
uncertainties and integrating over this PDF. It has been noted that averaging
over systematic uncertainties in itself is a Bayesian approach~\cite{Cousins:1992a} . Therefore the method presented is refered to as ``semi-Bayesian'', combining classical and Bayesian elements. We return to this point in section~\ref{sec:cov}.
Usually, uncertainties are assumed to be
described by a Gaussian distribution, which we will adopt for the remainder of 
this  paper. The implementation, however, makes it easy to use other parametrizations for the uncertainties.

We will refer in the following to the parameters with systematic uncertainties also as {\it nuisance parameters}.\par 
Two examples of how the PDF modifies if systematic uncertainties are
present are the following. 
In the case that the only uncertainty present is a theoretical uncertainty of the
background process the PDF is modified to:
\begin{equation}
q(n)_{s+b} = \frac{1}{\sqrt{2\pi}\sigma_{b}}\intop_0^{\infty}p(n)_{s+b'}\;\;e^{-\frac{(b-b')^2}{2\sigma_b^2}}\,db'
\label{eq:b}
\end{equation}
 Here $b$ is the estimated background level, 
and $\sigma_b$ is the uncertainty in the background estimation. 
If, in addition to the theoretical uncertainty for background, there
is the need to include the uncertainty in the signal detection efficiency the expression for $q(n)_{s+b}$ might be
extended to:
\begin{equation}
\begin{array}{ll}
q(n)_{s+b} =  
\frac{1}{2\pi\sigma_{b}\sigma_{\epsilon}} \times & \\
\intop_0^{\infty}\intop_0^{\infty}p(n)_{b'+ 
\epsilon's}\;\;e^{\frac{-(b-b')^2}{2\sigma_b^2}}\;\;
e^{\frac{- (1 -\epsilon')^2}{2\sigma_{\epsilon}^2}}db'd\epsilon' &\\
\end{array}
\label{eq:epsilon}
\end{equation}
where $\sigma_{\epsilon}$ is the uncertainty in the detection
efficiency expressed in {\em relative\/} terms with respect to 
the nominal efficiency. 
It is important to realize that the integration variables, here $\epsilon'$ and $b'$, are the possible ``true'' (but unknown) values of nuisance parameter. This indicates that this method is based on Bayesian statistics.

\section{\label{sec:cos} \texttt{\large POLE}: A General Algorithm for Confidence Belt Construction}
The integrals (\ref{eq:b}) and (\ref{eq:epsilon}) can be solved using
different methods. We note, however, that they are examples of 
simplified cases. The most general experimental situation involves both an 
uncertainty in signal efficiency as well as in the background 
detection efficiency, which are usually correlated, and possibly an
additional theoretical uncertainty in the background process prediction.
We have developed an algorithm that takes these effects into
account with the proper correlations between them. The algorithm
performs a Monte Carlo integration over the systematic uncertainties.
It has been implemented as a {\texttt{FORTRAN}} program,
\texttt{POLE} (POissonian Limit Estimator)~\cite{Conrad:2001a}. 
In the examples used in this section a Gaussian distribution of 
uncertainties is assumed, but the algorithm makes it easy to implement 
PDFs other than Gaussian (see the next section for an example of using
a different distribution). For the moment the code supports a Gaussian, flat and log-normal parametrization of the uncertainties.  
After determining the PDF through evaluation of the integrals,
different ordering schemes can be applied for the final calculation of the confidence belt. 
The results presented here are mainly for the likelihood ratio ordering scheme
with and without conditioning. We restrict ourselves to present
systematic uncertainties of signal and background efficiencies
separately to give a clear idea of the effect of varying a single
variable at a time. Real applications usually combine those uncertainties.\par

\begin{figure}[t]
\begin{minipage}[t]{0.92\linewidth}
\epsfig{file=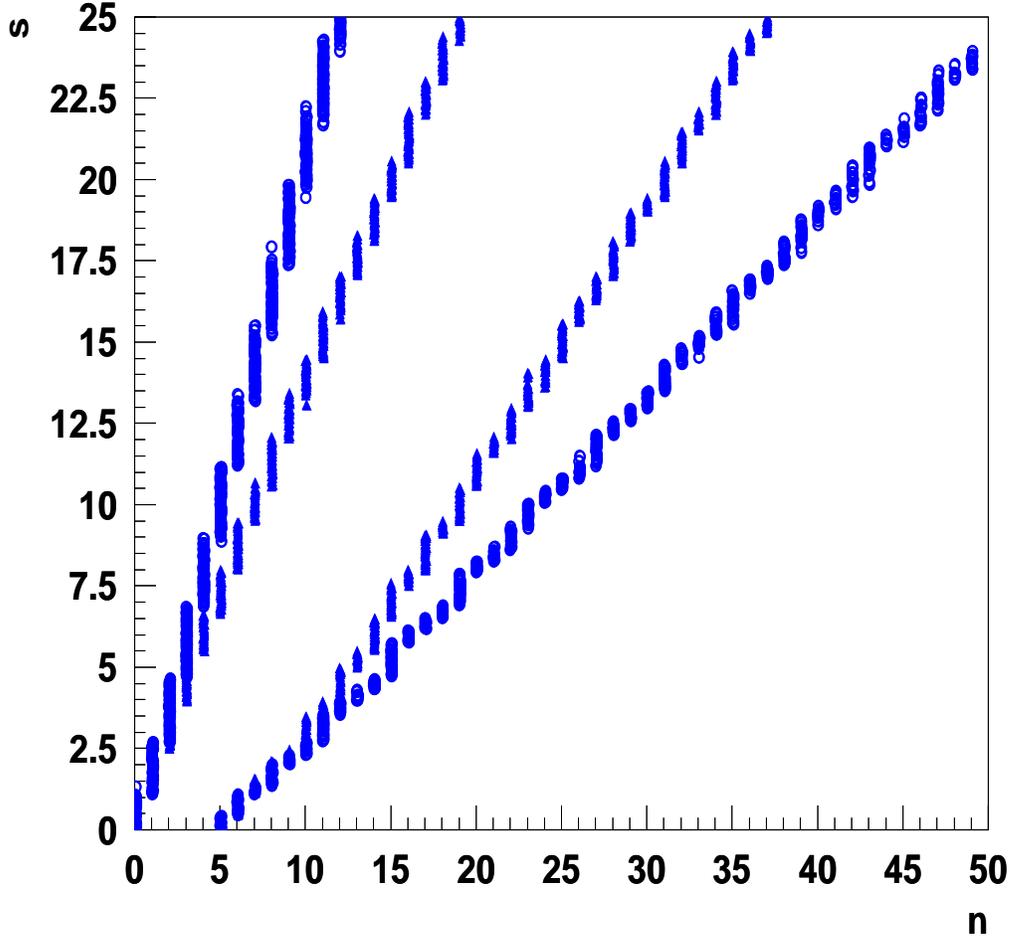,height=\linewidth,width=\linewidth}
\caption{ 90\% confidence belts obtained with \texttt{POLE} using the likelihood ratio ordering scheme and assuming different uncertainties in the signal
efficiency. The inner band has been constructed assuming no
uncertainty in the signal efficiency. The outer band  represents the
belt constructed with a signal efficiency uncertainty of 40\%. The background
expectation in this particular case was b = 2}
\label{fig:a}
\end{minipage}\hfill
\end{figure}

The confidence belt constructions have been performed using steps of 0.05 in signal expectation and performing the construction up to a maximal signal expectation
of 50 and a maximal number of detected events of 100.
Including systematic uncertainties generally leads to a widening 
of the confidence belt. Figure~\ref{fig:a} shows an example of a
likelihood ratio confidence belt construction  with and without
uncertainty in the signal efficiency, where a background expectation
$b= 2$ has been assumed.

\begin{figure*}[t]
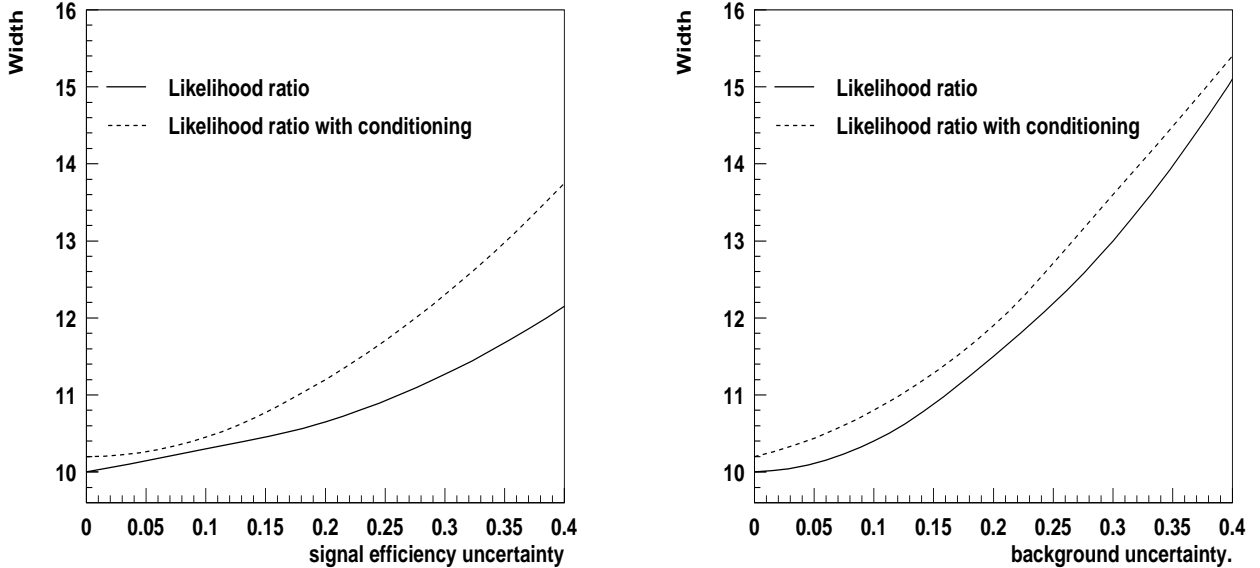

\begin{minipage}[t]{0.46\linewidth}
\centering\epsfig{file=fig3a.epsi,height=\linewidth,width=\linewidth}
\end{minipage}\hfill
\begin{minipage}[t]{0.46\linewidth}
\centering\epsfig{file=fig3b.epsi,height=\linewidth,width=\linewidth}
\end{minipage}
\caption{Example of the dependence of the likelihood ratio
confidence interval width on the systematic uncertainties, with and
without conditioning, as obtained with \texttt{POLE}. The left plot shows the width as a
function of the uncertainty in signal efficiency assuming no
additional uncertainty in background expectation. The right plot shows 
the width as a function of the background uncertainty. We have used 
$n_o$=17 and a background of 15 in constructing the plots.}
\label{fig:11}
\end{figure*}

\begin{figure*}[t]
\epsfig{file=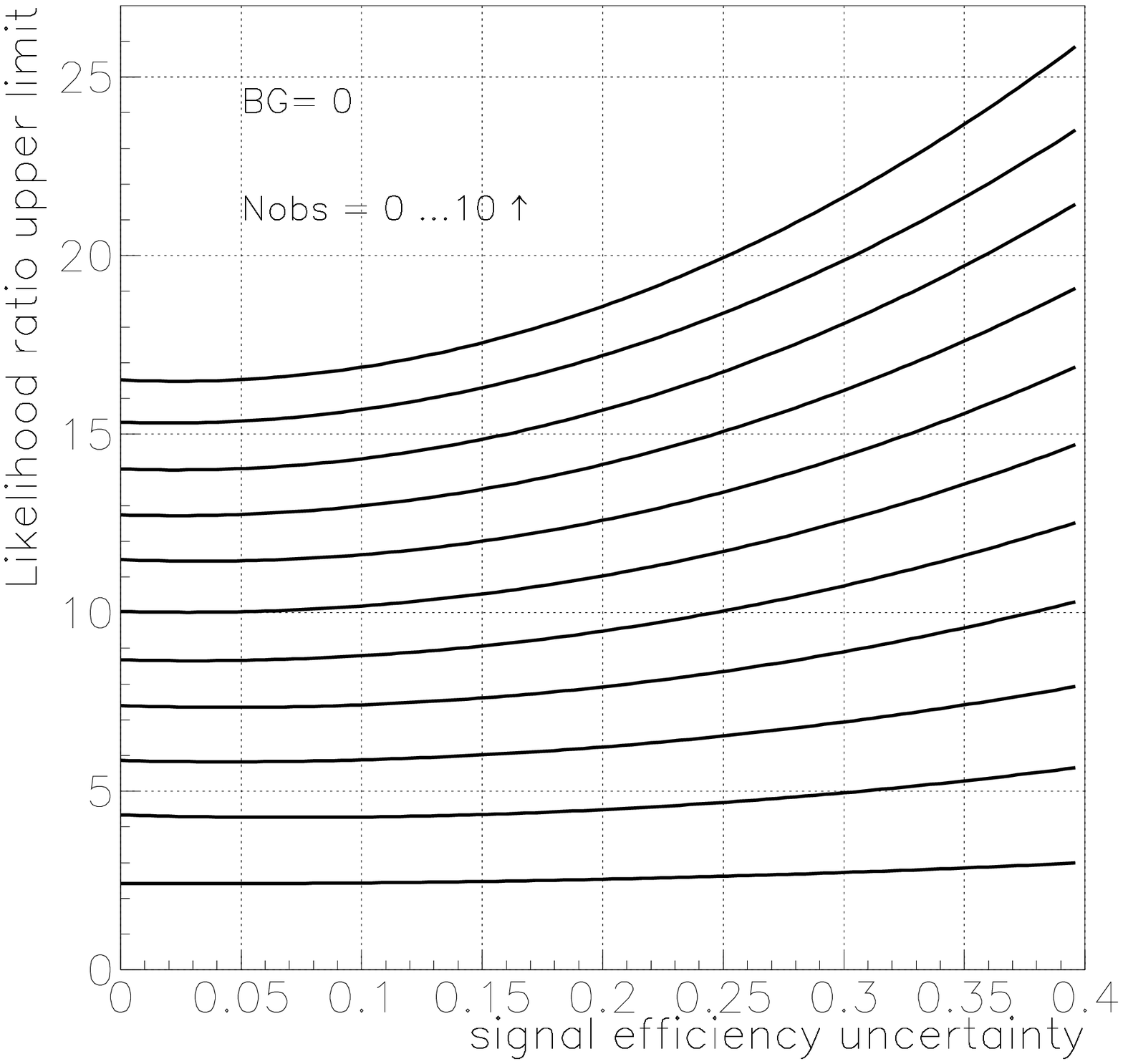,height=7cm,width=7cm}
\epsfig{file=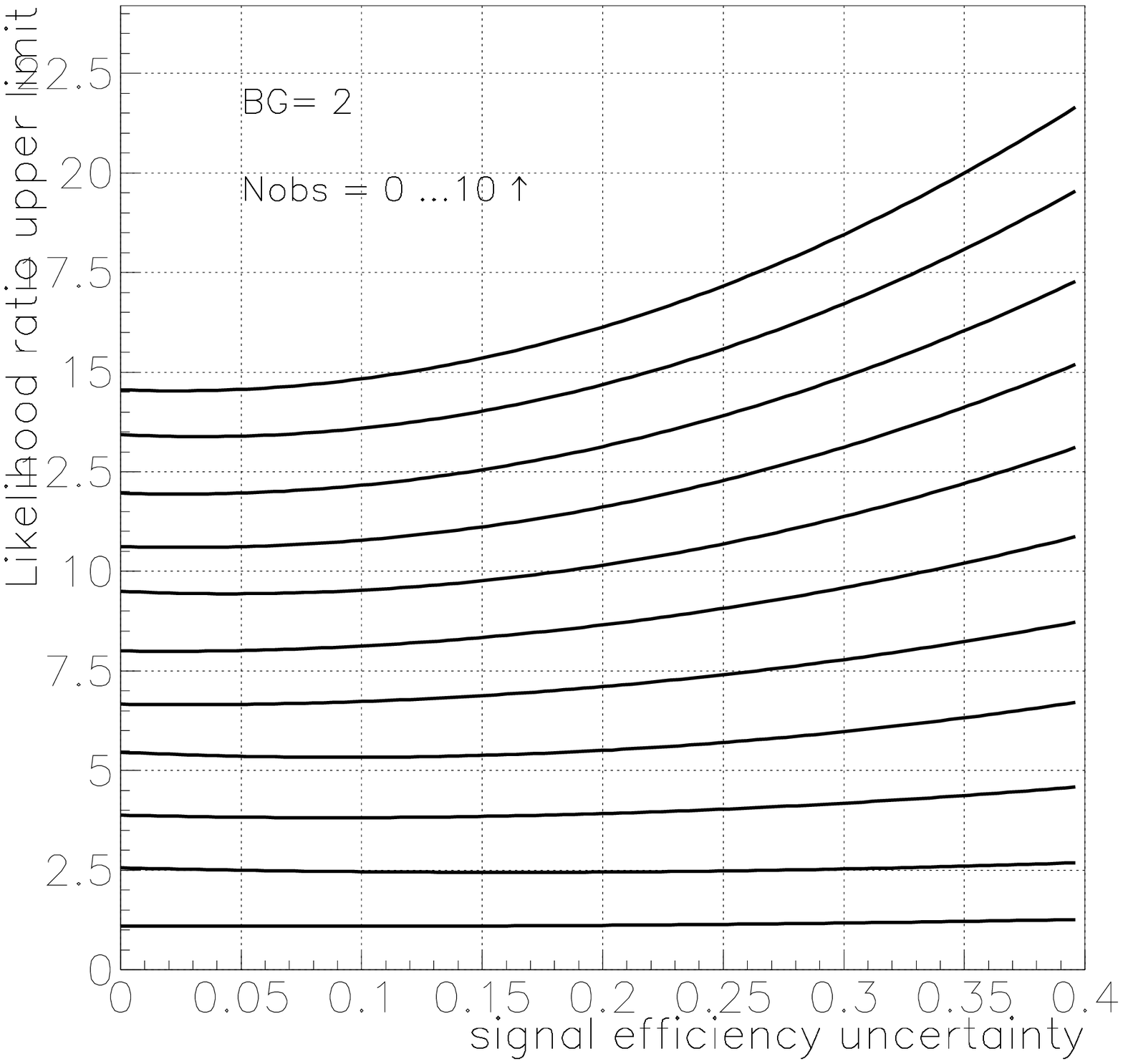,height=7cm,width=7cm}
\epsfig{file=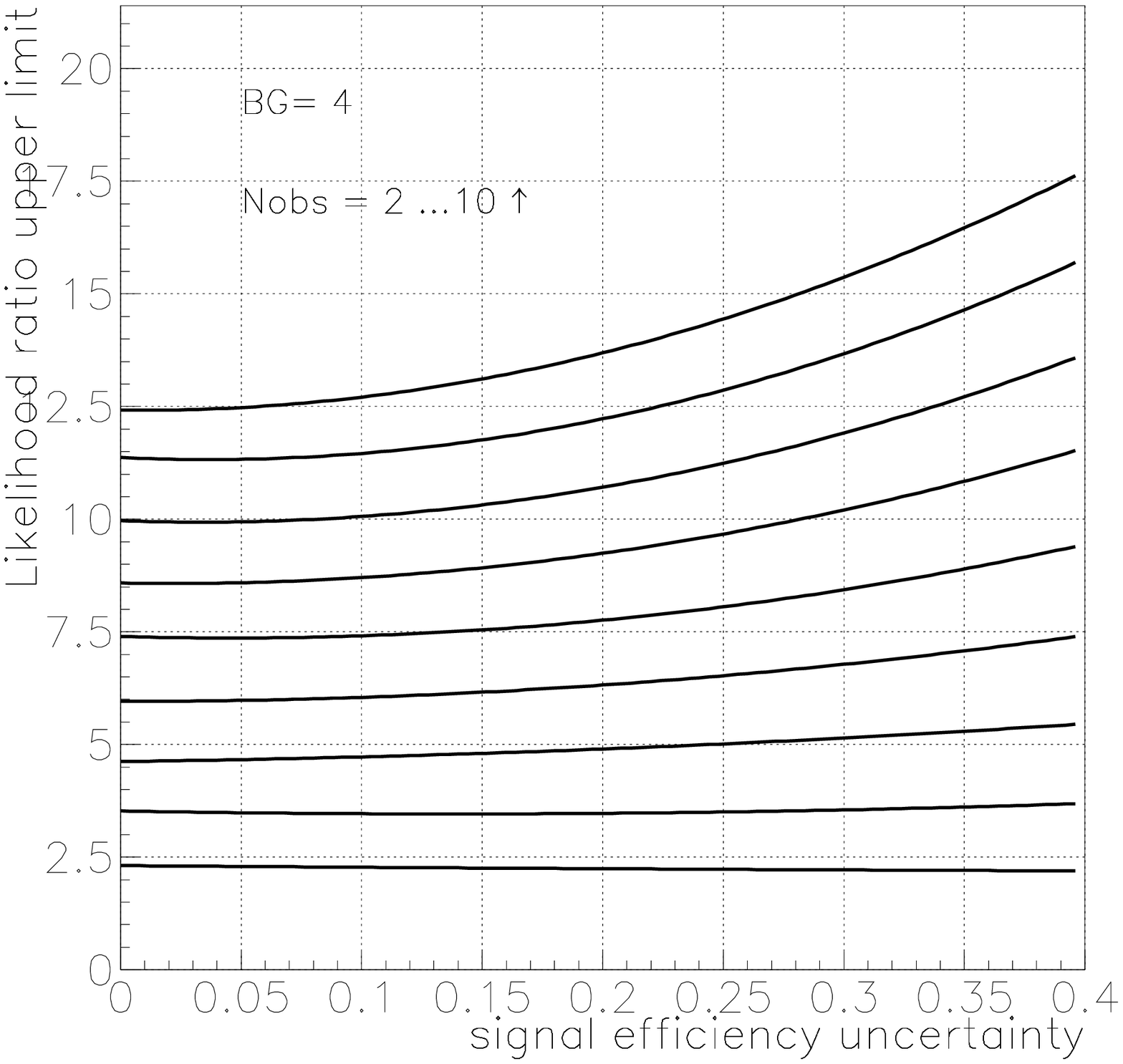,height=7cm,width=7cm}
\epsfig{file=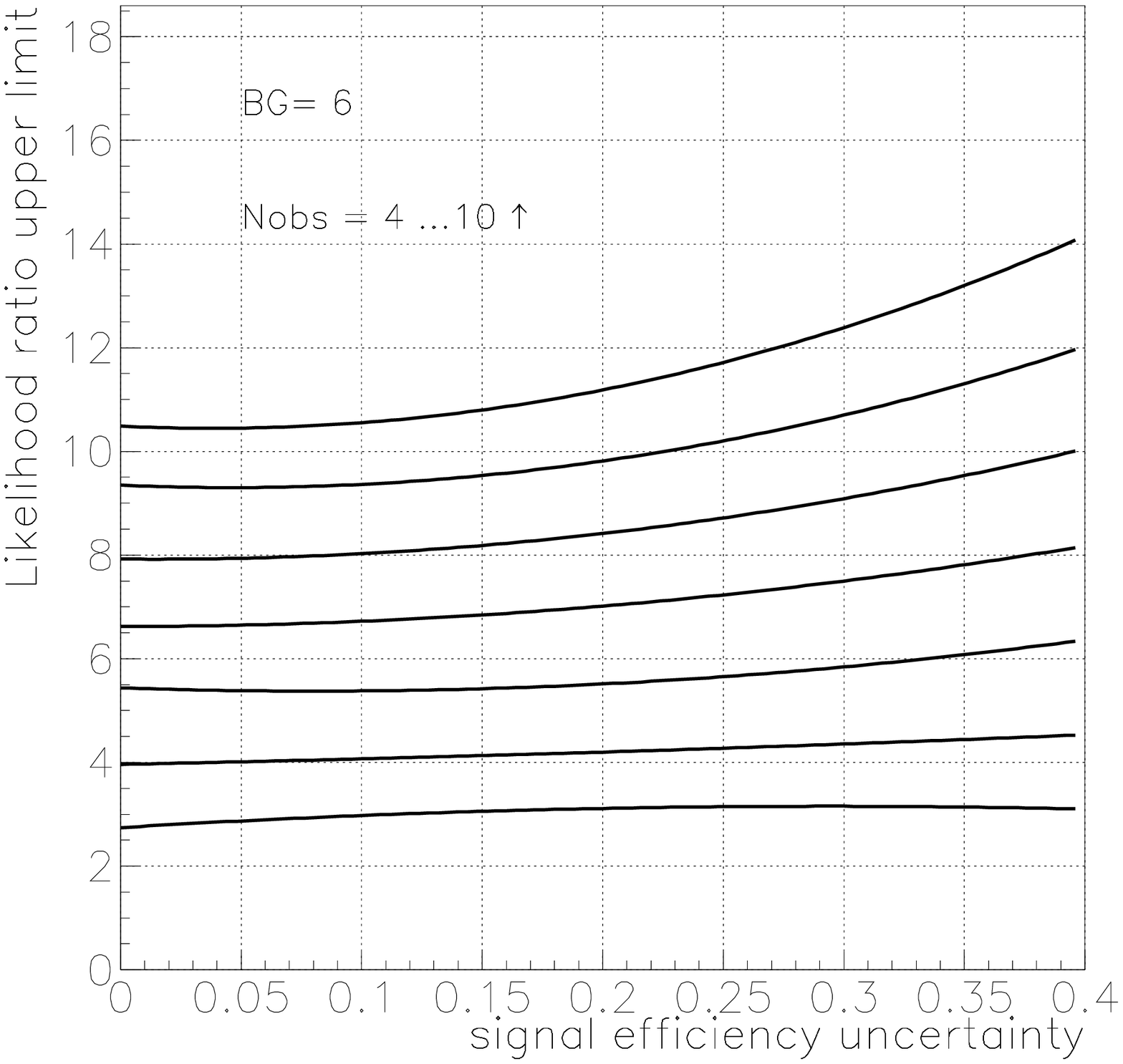,height=7cm,width=7cm}
\epsfig{file=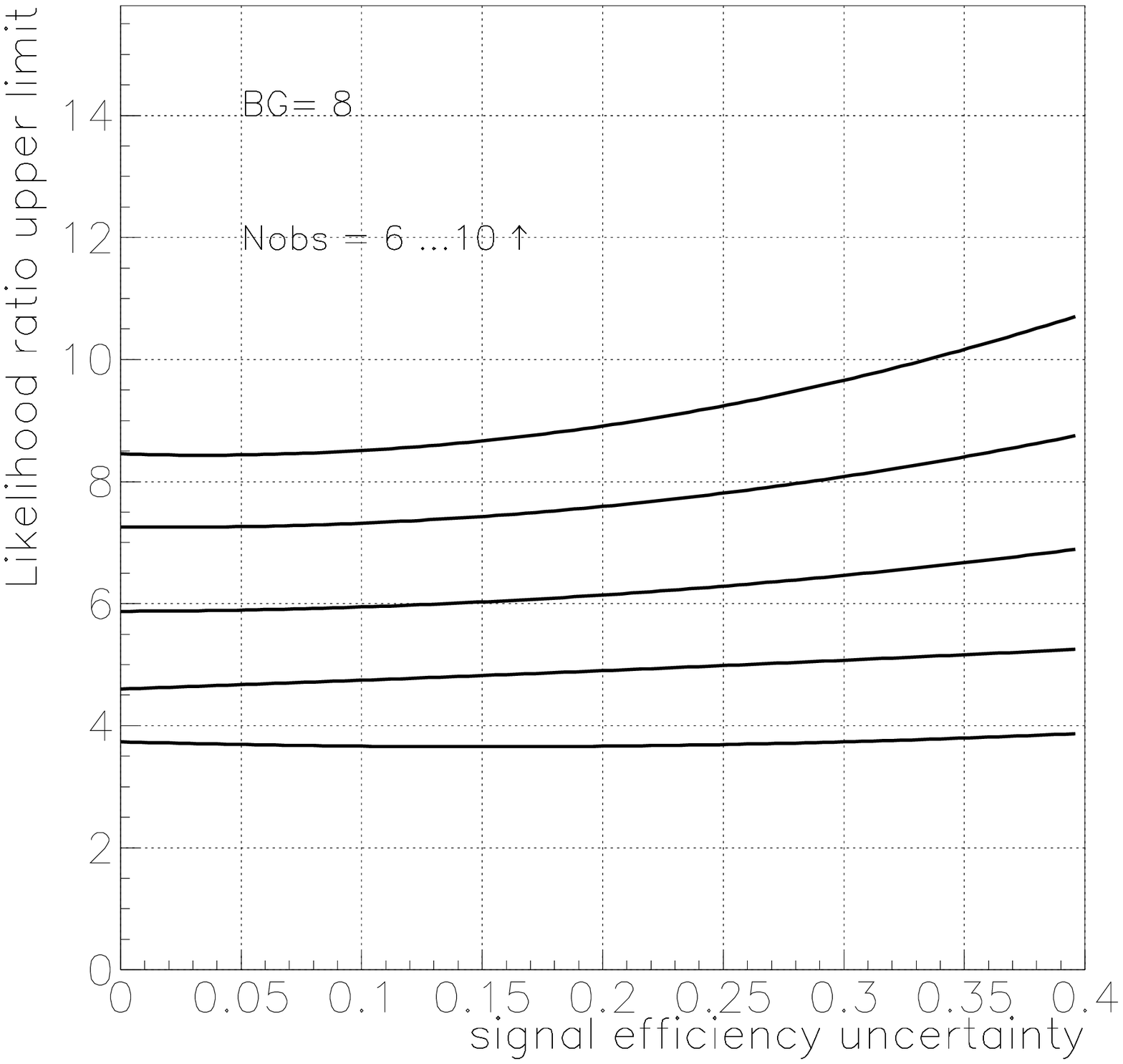,height=7cm,width=7cm}
\epsfig{file=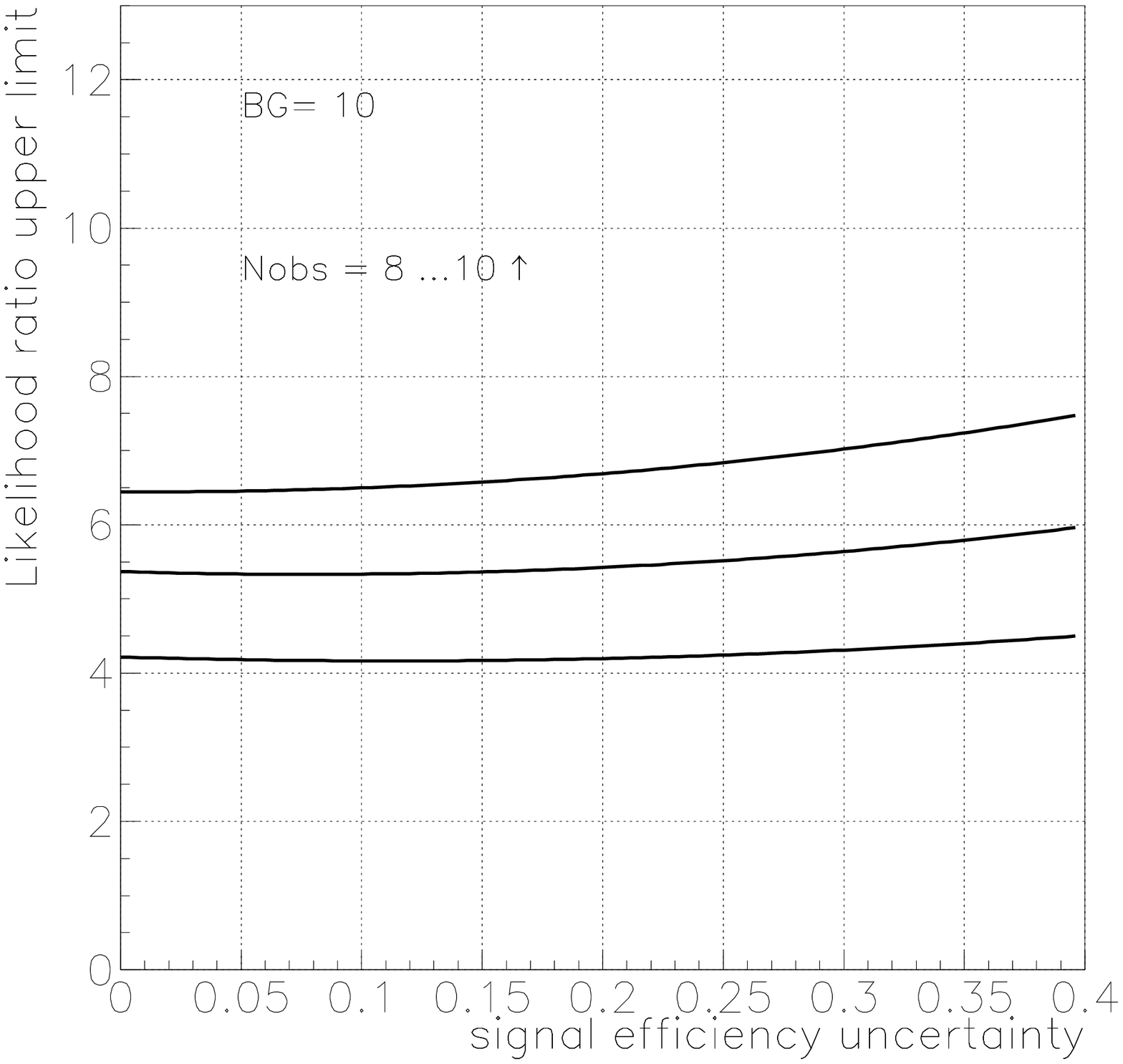,height=7cm,width=7cm}
\caption{Likelihood ratio upper limit as a function of signal efficiency uncertainty for expected backgrounds = 0, 2, 4, 6. 8 and 10. Cases with number of observed events significantly less than expected background have been omitted.}
\label{fig:extended}
\end{figure*}

\begin{figure*}[t]
\epsfig{file=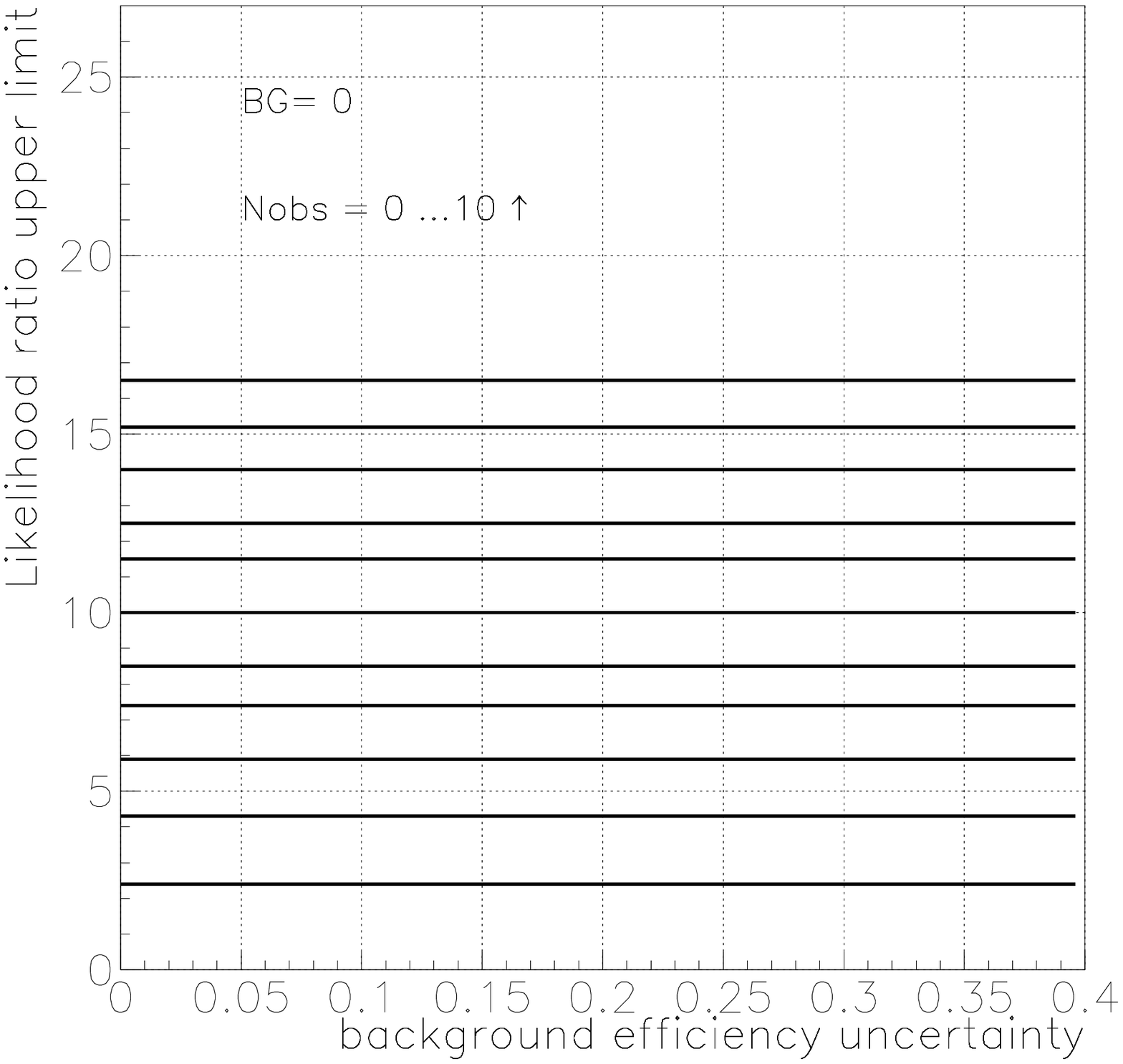,height=7cm,width=7cm}
\epsfig{file=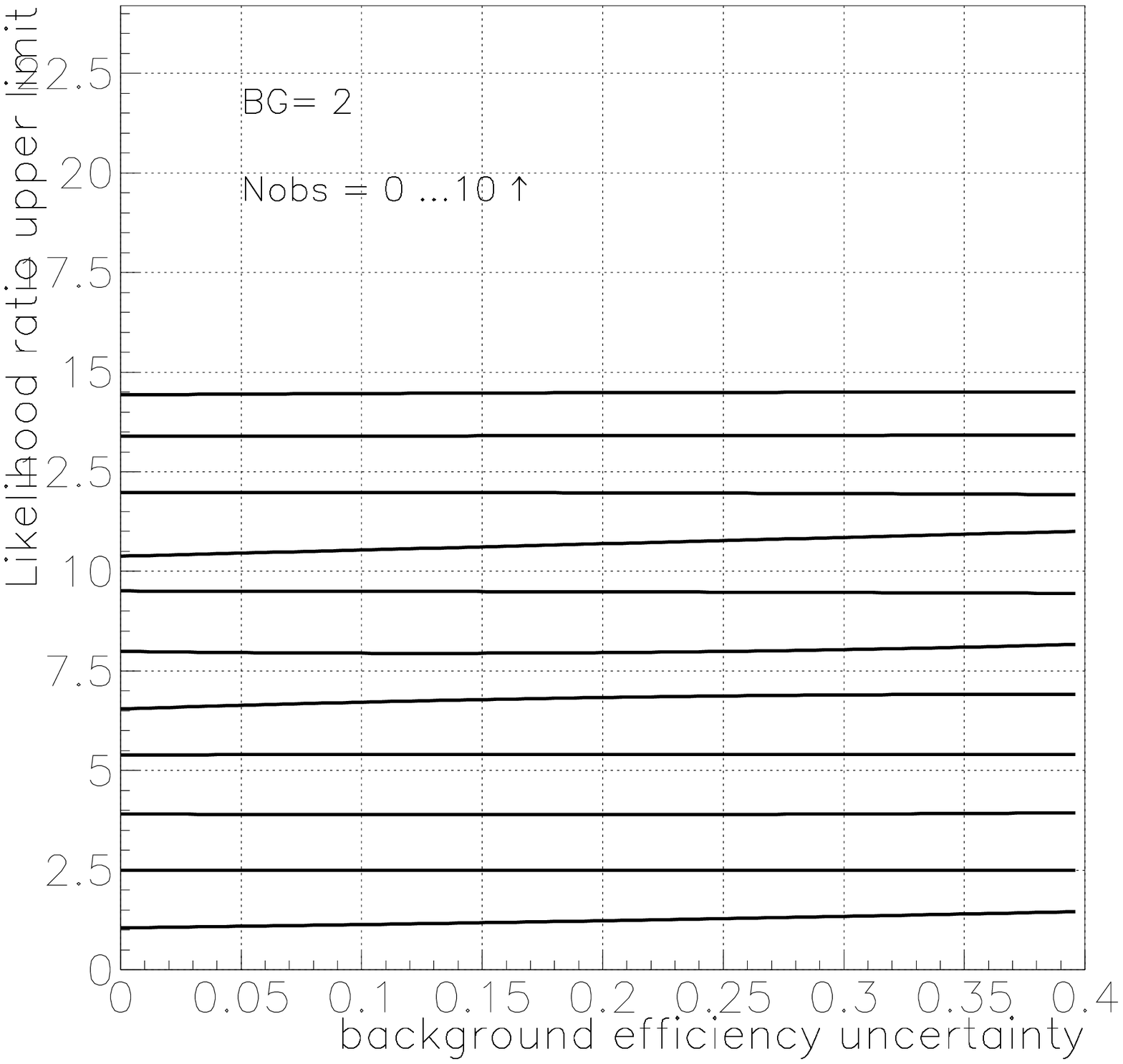,height=7cm,width=7cm}
\epsfig{file=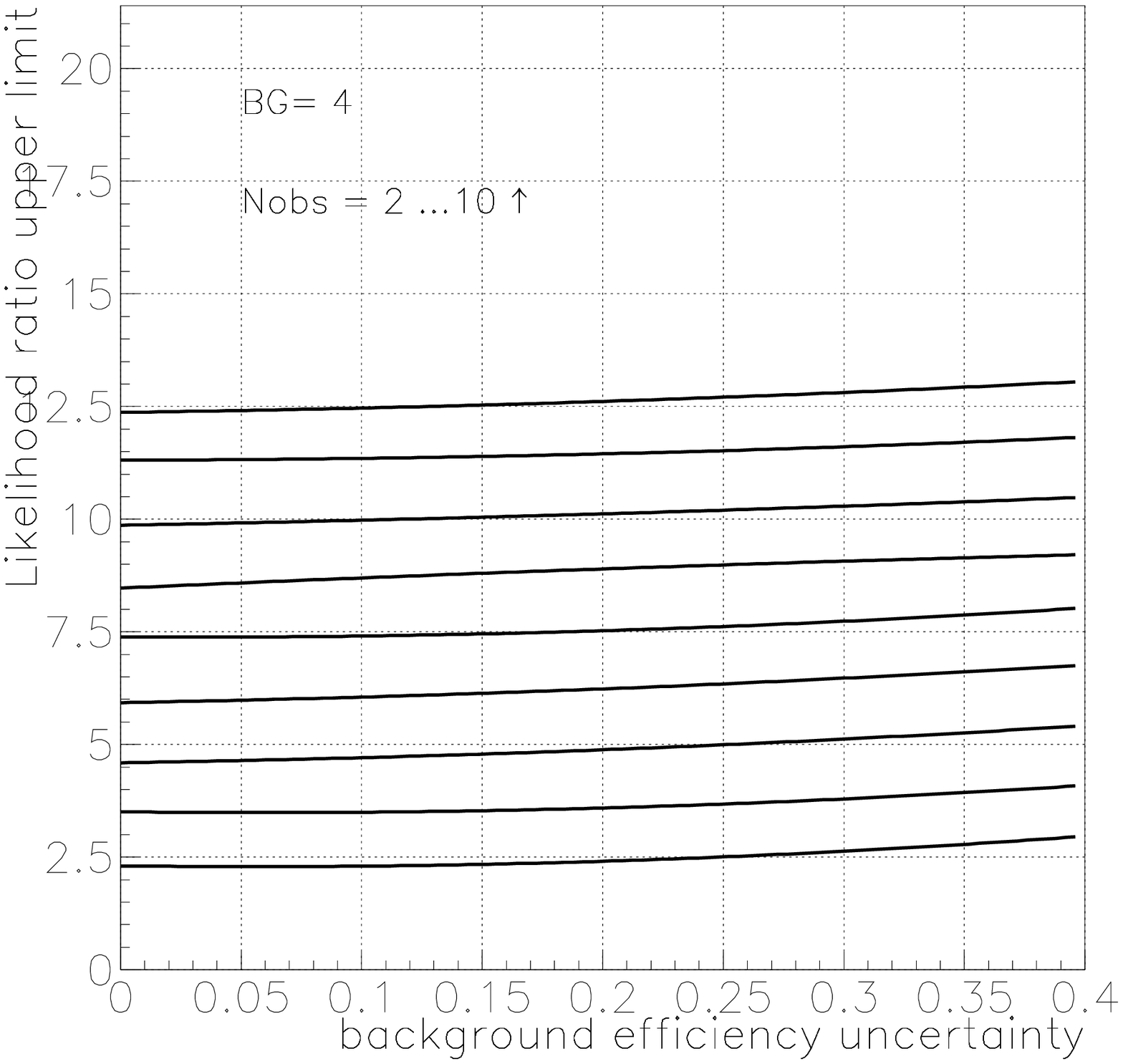,height=7cm,width=7cm}
\epsfig{file=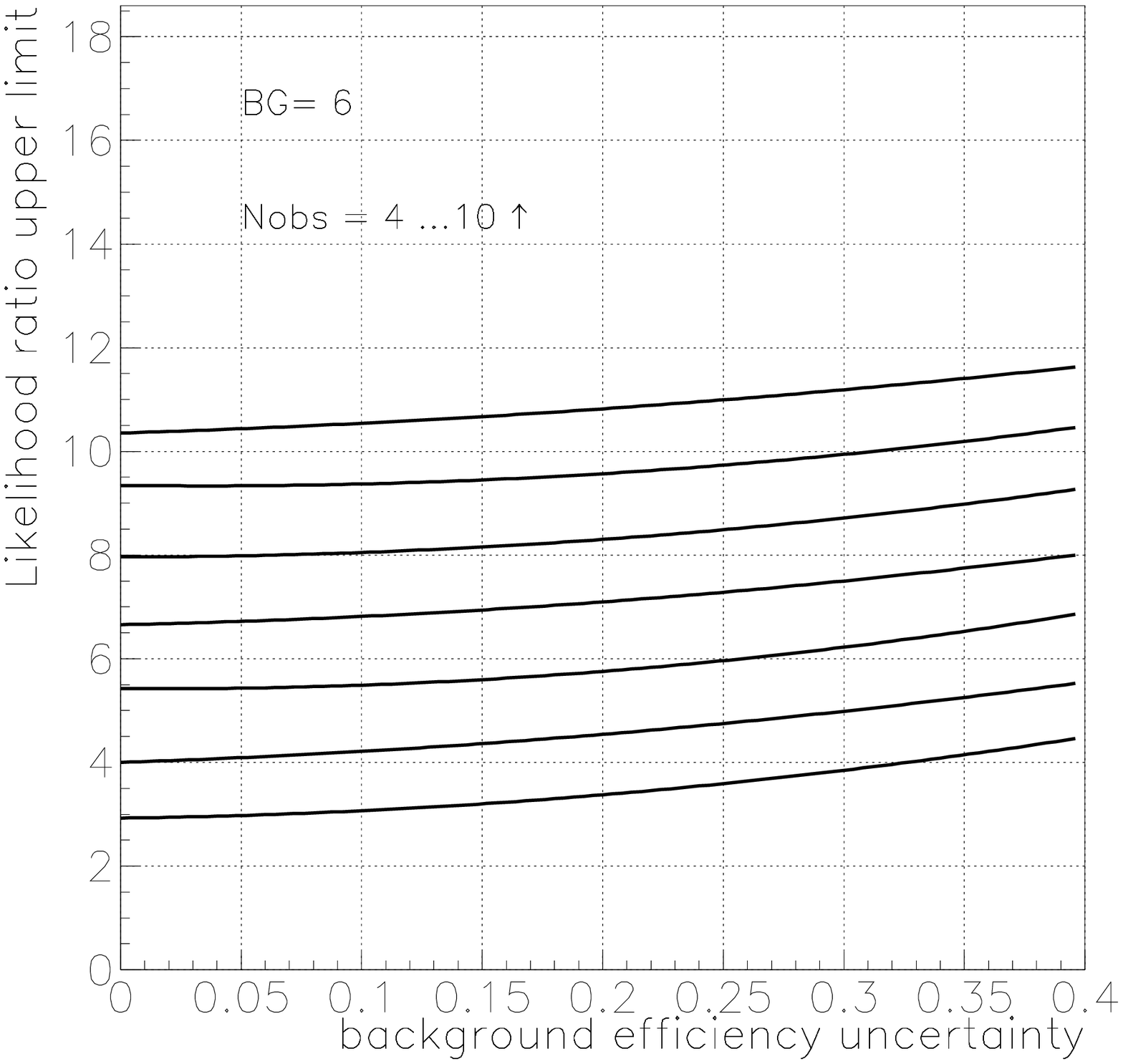,height=7cm,width=7cm}
\epsfig{file=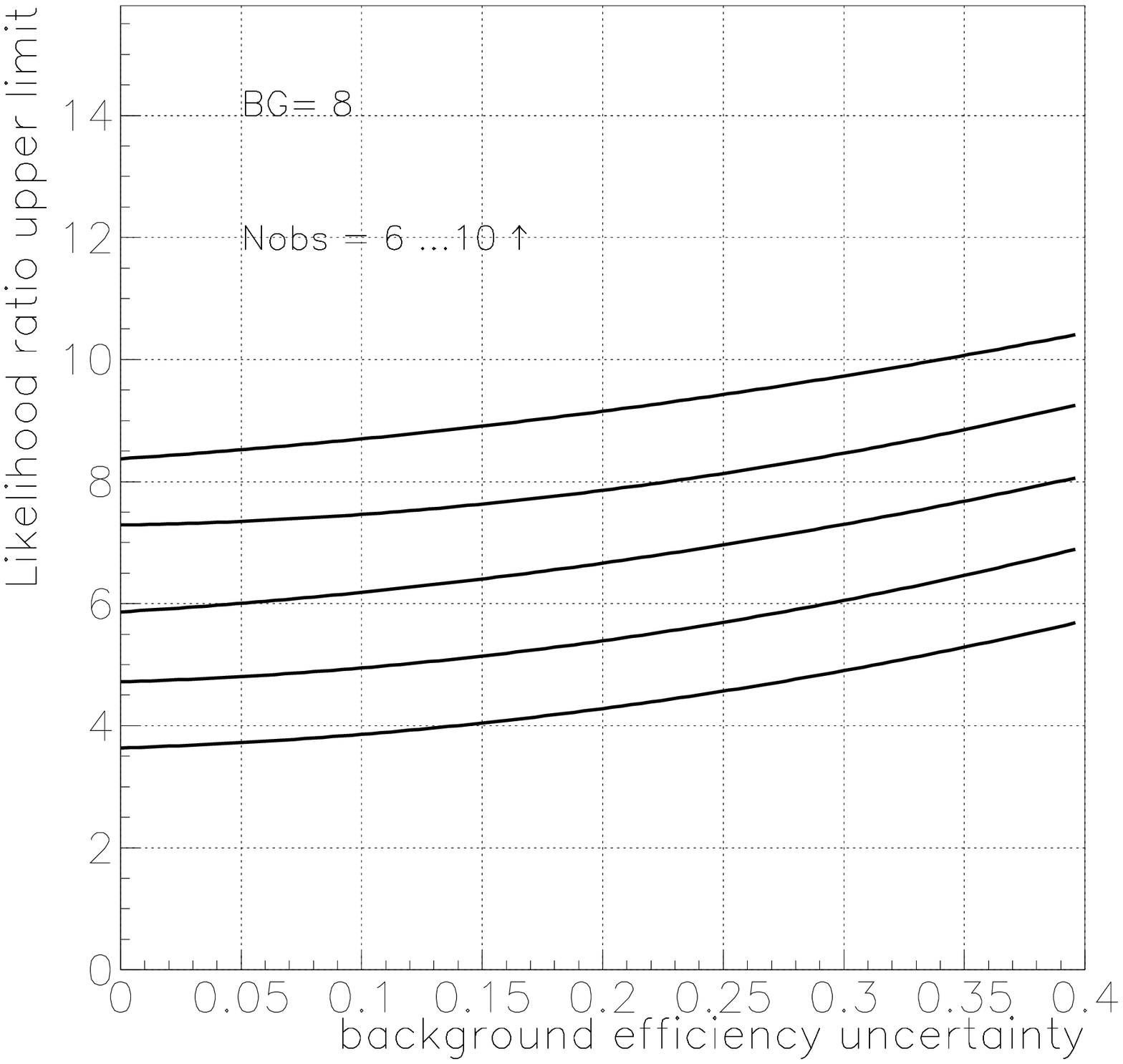,height=7cm,width=7cm}
\epsfig{file=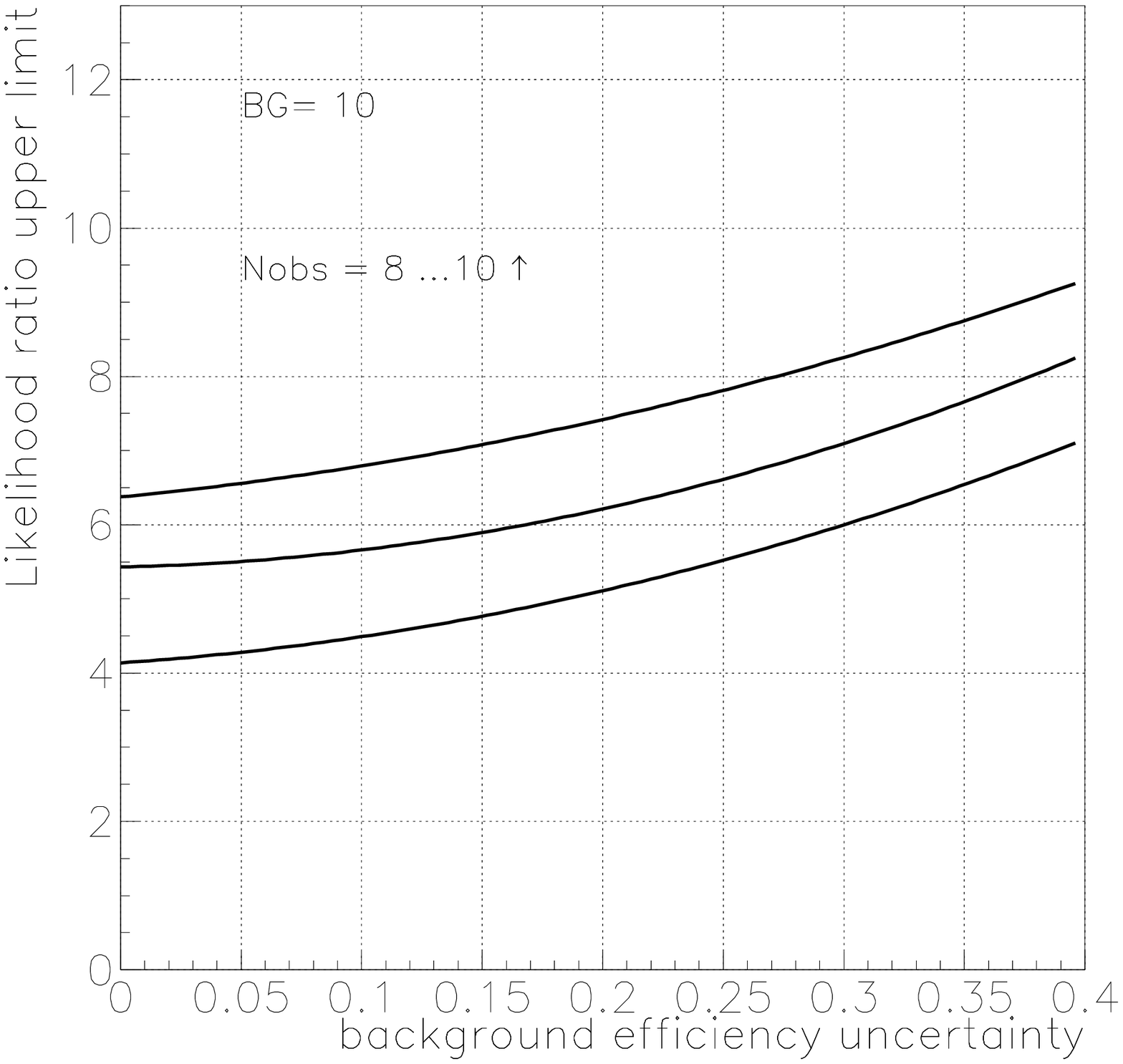,height=7cm,width=7cm}
\caption{Likelihood ratio upper limit as a function of background efficiency uncertainty for expected backgrounds = 0, 2, 4, 6. 8 and 10. Cases with number of observed events significantly less than expected background have been omitted.}
\label{fig:extended2}
\end{figure*}

Examples for some resulting intervals are given in tables~\ref{tab:sys} and~\ref{tab:sys2}. 
Different combinations of number of observed events, $n_0$, and
expected background, $b$, are given for different uncertainties in the
signal and background efficiency. \\
The width of the interval for two particular
examples of observed events and expected background as function of
signal efficiency uncertainty and background uncertainty 
is shown in figure \ref{fig:11}. Note that for low background
expectation, the  uncertainties in the background can be neglected
(see also table~\ref{tab:sys2}). Figures \ref{fig:extended}  and \ref{fig:extended2} give more extended information on resulting intervals.

\begin{table}[t]
\begin{center}
\begin{tabular}{|l|l|l|l|l|}
\hline\hline
$n_0$ 	&$b$ &signal efficiency & Likelihood Ratio  &Likelihood Ratio interval \vspace{-0.1cm}\\
	&    & uncertainty  (\%)  &interval       & with conditioning\\ \hline
2     	&2       &0               &0: 3.90     	&0: 4.00\\
     	&        &0.2             &0: 3.95       &0: 4.34         \\
      	&        &0.3             &0: 4.10	&0: 4.75              \\
      	&        &0.4             &0: 4.65	&0: 5.35        \\ \hline
3     	&2       &0               &0: 5.40      &0: 5.30          \\
      	&        &0.2             &0: 5.70	&0: 5.65         \\
      	&        &0.3             &0: 5.95	&0: 6.20 \\
      	&        &0.4             &0: 6.80	&0: 7.10  \\ \hline
4     	&2       &0               &0: 6.60      &0: 6.60   \\
      	&        &0.2             &0: 7.10	&0: 7.30   \\
      	&        &0.3             &0: 7.75	&0: 7.85  \\
      	&        &0.4             &0: 8.95	&0: 9.15  \\ \hline
5       & 2      &0               &0.40: 7.95     &0.50: 8.05\\
        &        &0.2             &0.40: 8.60      &0.50: 8.60 \\
        &        &0.3             &0.40: 9.55     &0.50: 9.65 \\
        &        &0.4             &0.40:11.15    &0.50:11.20 \\ \hline
6       & 2      &0               &1.10: 9.45    &1.10: 9.45 \\
        &        &0.2             &1.05,10.05   &1.05:10.10 \\
        &        &0.3             &1.05:11.50   &1.05:11.50\\
        &        &0.4             &1.05:13.35   &1.05:13.35\\  \hline\hline   
\end{tabular}
\caption{\label{tab:sys}Examples of likelihood ratio 90\% confidence intervals including systematic
uncertainty in the signal efficiency and assuming no uncertainty in
the background prediction.}
\end{center}
\end{table}

\begin{table}[t]
\begin{center}
\begin{tabular}{|l|l|l|l|l|}
 \hline \hline
$n_0$ &$b$ &background & Likelihood ratio     &Likelihood ratio \vspace{-0.1cm}\\ 
      &   &uncertainty & interval             &interval with conditioning\\ \hline
2     &2       &0         &0: 3.90 	  & 0: 4.00 \\
      &        &0.2       &0: 3.95        & 0: 4.10  \\
      &        &0.3       &0: 3.95        & 0: 4.25  \\
      &        &0.4       &0: 3.95        & 0: 4.35  \\ \hline
3     &2       &0         &0: 5.40        & 0: 5.30   \\
      &        &0.2       &0: 5.45        & 0: 5.35  \\
      &        &0.3       &0: 5.45        & 0: 5.45  \\
      &        &0.4       &0: 5.50        & 0: 5.55  \\ \hline
4     &2       &0         &0: 6.60        & 0: 6.60  \\
      &        &0.2       &0: 6.95        & 0: 6.65  \\
      &        &0.3       &0: 6.95        & 0: 6.80  \\
      &        &0.4       &0: 6.95        & 0: 6.80  \\ \hline
5     &2       &0         &0.40: 7.95      & 0.50: 8.05          \\
      &        &0.2       &0.35: 7.95     &  0.50: 8.10 \\
      &        &0.3       &0.30: 8.00      & 0.50: 8.10  \\ 
      &        &0.4       &0.20: 8.20      & 0.45: 8.15\\ \hline
6     &2       &0.        &1.10: 9.45      & 1.10: 9.45\\ 
      &        &0.2       &1.05: 9.45      & 1.10: 9.50\\
      &        &0.3       &1.00: 9.50        & 1.05: 9.50 \\
      &        &0.4       &0.95: 9.50       & 1.00: 9.50\\ \hline
 \hline \hline
\end{tabular}
\caption{\label{tab:sys2}Examples of likelihood ratio 90\% confidence intervals including systematic
uncertainty in the background expectation and assuming no uncertainty
in the signal efficiency.}
\end{center}

\end{table}

An interesting case arises when there are significantly less events observed 
than expected from background and there is an uncertainty in the
signal efficiency. In this case, the width of the confidence interval does not increase (see table~\ref{tab:sys3}). 
Note that if we use conditioning the effect disappears.
The same can not be observed in the case where we only consider an increasing
background uncertainty (see table~\ref{tab:sys4}).\\  
\begin{table}[t]
\begin{center}
\begin{tabular}{|l|l|l|l|l|}
\hline\hline
$n_0$ &$b$ &signal efficiency  & Likelihood ratio &Likelihood ratio\vspace{-0.1cm}\\ 
      &    & uncertainty      & interval &interval with conditioning\\ \hline
2     &6       &0               &0: 1.55      &0: 3.15\\
      &        &0.2             &0: 1.55      &0: 3.35\\
      &        &0.4             &0: 1.45      &0: 4.00\\ \hline
4     &6       &0               &0: 2.85      &0: 4.30 \\
      &        &0.2             &0: 3.20      &0: 4.60 \\
      &        &0.4             &0: 3.35      &0: 5.35\\\hline\hline
\end{tabular}
\caption{\label{tab:sys3} Likelihood Ratio confidence intervals with systematic
uncertainty in the signal efficiency and no uncertainty in the
background expectation. Here two examples are shown where there are less events
observed than expected background:
The interval does not increase with increasing uncertainty if there are
significantly less events observed than expected background. However,
if expected background and number of observed events are comparable,
the interval becomes larger. In case conditioning is applied, it grows
larger in all cases.}
\end{center}
\end{table}

\begin{table}[t]
\begin{center}
\begin{tabular}{|l|l|l|l|l|}
\hline\hline
$n_0$ &$b$ &background   &Likelihood Ratio  &Likelihood ratio\vspace{-0.1cm}\\      &    &uncertainty (\%) &interval       &interval with conditioning\\ \hline
2     &6       &0               &0: 1.55       &0: 3.15              \\
      &        &0.2             &0: 1.55       &0: 3.50             \\
      &        &0.4             &0: 2.64       &0: 3.85
\\ \hline 
4     &6       &0               &0: 2.85       &0: 4.30  \\           
      &        &0.2             &0: 3.25       &0: 4.55  \\
      &        &0.4             &0: 4.60       &0: 5.55   \\
\hline\hline
\end{tabular}

\caption{\label{tab:sys4} Likelihood ratio  confidence intervals with systematic uncertainty in the background expectation and no uncertainty in the
signal efficiency. Here two examples are shown where there are less events
observed than expected background: The confidence interval becomes larger with increasing
uncertainty in the background expectation.}
\end{center}
\end{table}

\subsection{Negative values of the nuisance parameters: Using a log - normal distribution.}
\begin{figure}[t]
\begin{minipage}[t]{0.92\linewidth}
\epsfig{file=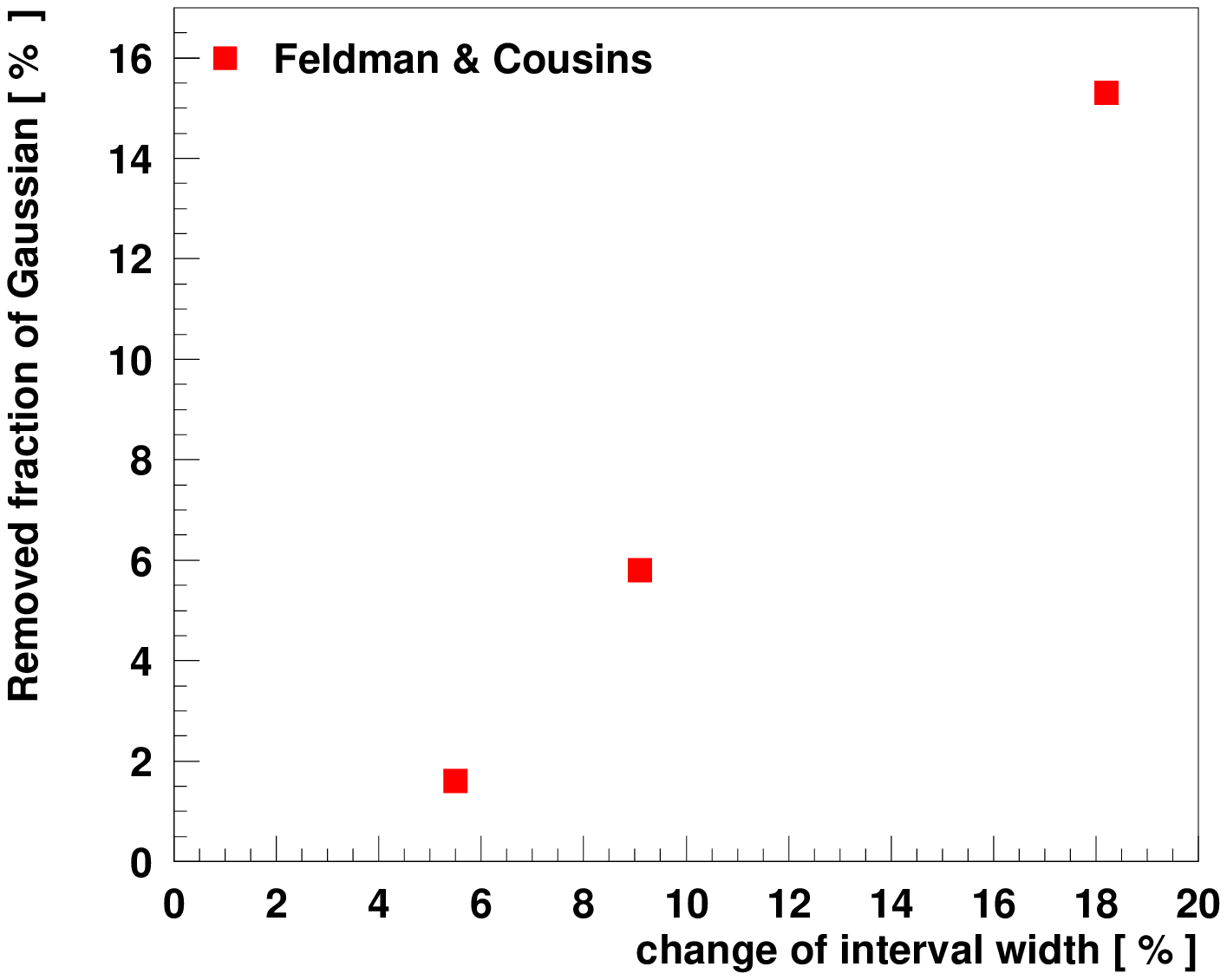,height=\linewidth,width=\linewidth}
\caption{The relative change of the likelihood ratio 
interval width as a function of fraction of Gaussian
removed. In this example, $n_o = 4$ and $b=4$ have been assumed. }
\label{fig::trunc}
\end{minipage}\hfill
\end{figure}

\begin{figure}[t]
\begin{minipage}[t]{0.92\linewidth}
\epsfig{file=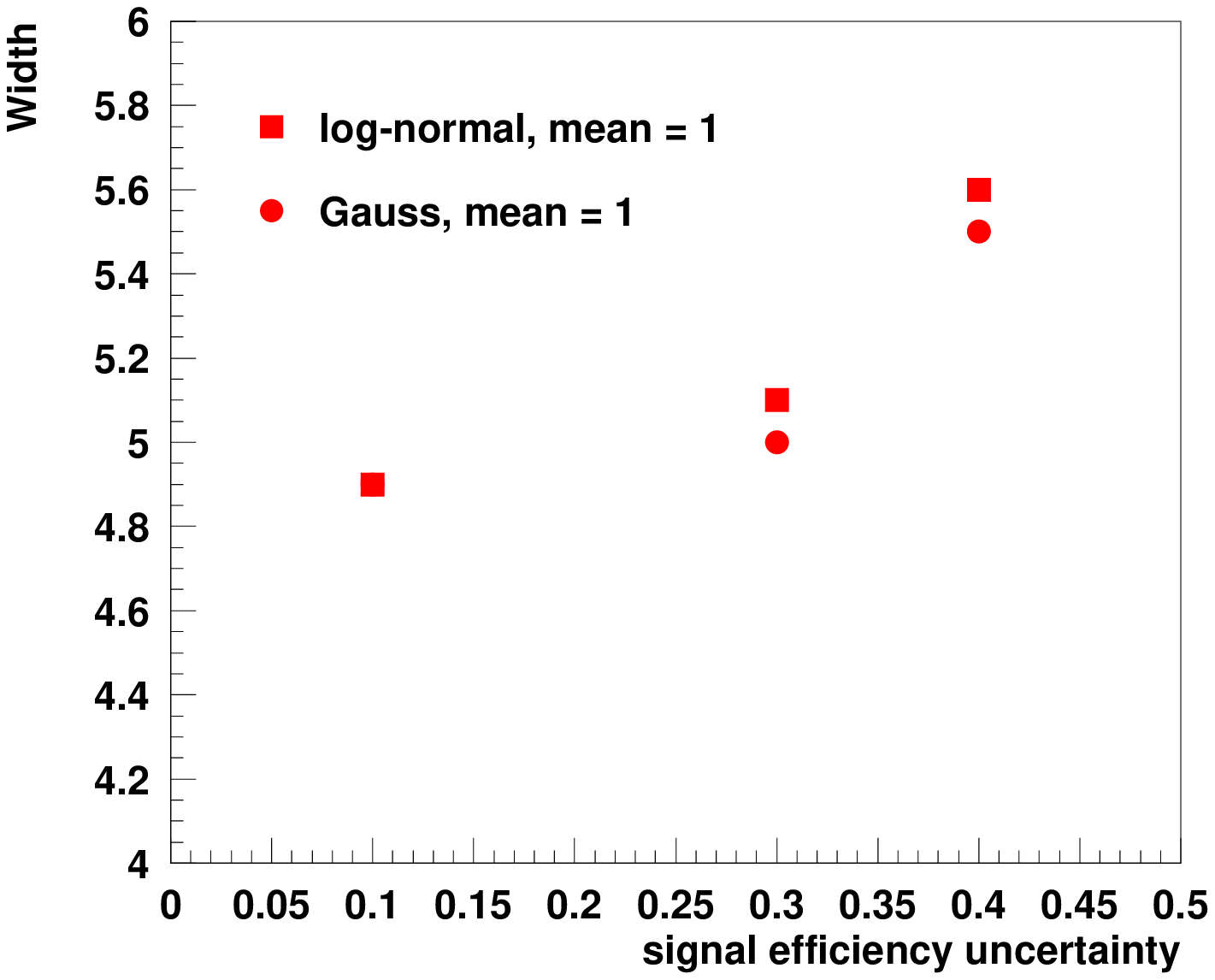,height=\linewidth,width=\linewidth}
\caption{Likelihood ratio confidence interval width as a function of signal efficiency
uncertainty for a Gaussian and a log-normal distribution with mean at 1. In this
example, $n_o = 4$ and $b=4$ have been assumed and the Gaussian was
truncated at zero.}
\label{fig::logN}
\end{minipage}\hfill
\end{figure}

In experimental situations where the systematic uncertainties are
high, a problem might arise due to the fact that sampling from a
Gaussian PDF allows negative values. \texttt{POLE} is dealing with these cases 
truncating the Gaussian distribution and renormalizing the part above
zero.\\
We examine the effect of truncating the Gauss distribution
by calculating the confidence interval for different values of
truncation point (see figure \ref{fig::trunc}). 
Considering a Gaussian distribution centered on one widh $\sigma$ = 40 \%, a truncation at zero removes only 0.7 \%. Figure \ref{fig::trunc} therefore indicates that effects on the confidence interval due to the truncation are 
negligible for all cases considered in this paper.\\
A PDF for the nuisance parameters extending to negative
values or which falls off to zero discontinuously is certainly
undesired from a conceptual point of view. We therefore test the
behavior of the confidence interval if we replace the Gaussian
distribution with a log-normal distribution, which in the general form
is given by:
\begin{equation}
q(x)_{\mu,\sigma} =  \frac{1}{\sqrt{2\pi}x
\sigma}\;e^{-\frac{(\ln{x}-\mu )^2}{2\sigma^2}}
\label{eq:logn}
\end{equation}
We require  the mean of
the log-normal distribution to be the nominal value of the nuisance parameter
and use the Gaussian standard deviation as before (the
variance of the log normal distribution will then be approximately the same).
The confidence interval for Neyman and likelihood ratio ordering
under these assumptions are shown for one particular example of number
of observed events and expected background as a function of signal
efficiency in figure \ref{fig::logN}.
The differences between using a Gaussian distribution and using a
log-normal distribution are generally small, in our example less than
$\sim$ 2~\%. The use of a log-normal distribution is implemented as an option in \texttt{POLE}. 

\section{\label{sec:comparison} Comparison with the $\chi^2$ method.}
Since the ratio of the likelihoods is asymptotically $\chi^2$ distributed the approximation:
\begin{equation}
\Delta\chi^2 = -2\times \min_\epsilon{\frac{\mathcal{L}(n_o,s,\epsilon)}{\mathcal{L}(n_o,s_{best},\epsilon_{best})}}
\end{equation}
is often used, see e.g. \cite{Feldman:2000}.\\
Here for a given observation,$n_0$, the $\Delta \chi^2 $ is calculated as a function of $s$ and a cut on $\Delta \chi^2$ is performed to obtain the confidence interval (e.g. $\Delta \chi^2$ = 2.71 corresponds to 90 \% confidence level for one degree of freedom).

\begin{figure*}[t]
\begin{minipage}[t]{0.92\linewidth}
\epsfig{file=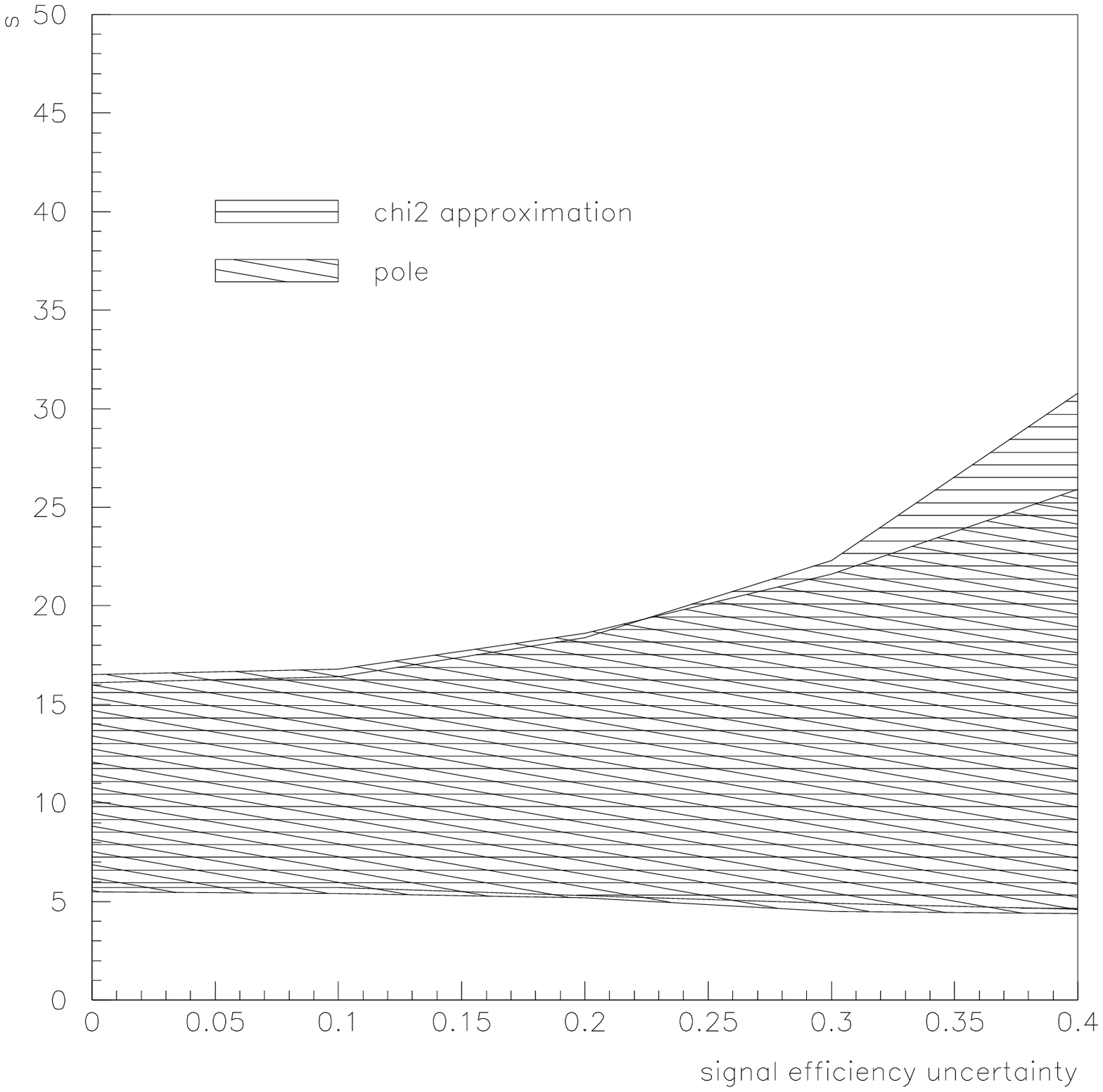,height=0.45\linewidth,width=0.45\linewidth}
\epsfig{file=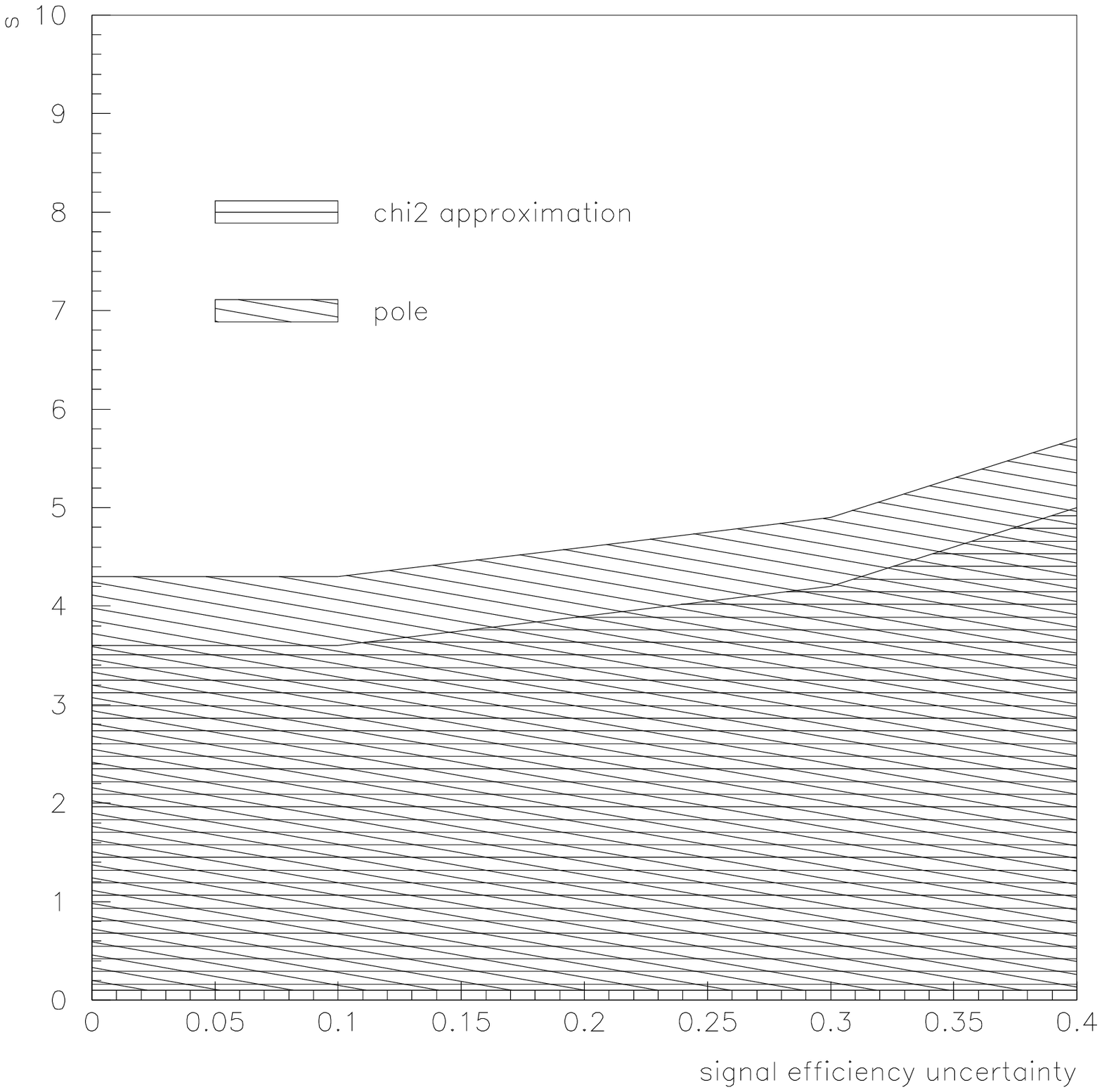,height=0.45\linewidth,width=0.45\linewidth}
\caption{The likelihood ratio confidence interval as calculated by {\tt pole} and using the $\chi^2$ approximation. Number of observed events is 10 (left panel) and 1 (right panel). The background was assumed to be zero in both cases.}
\label{fig:comparison}
\end{minipage}\hfill
\end{figure*}

Figure  \ref{fig:comparison} illustrates the effect of including uncertainties on the resulting confidence intervals for the $\chi^2$ approximation as compared to the method proposed here.\\
Generally, the $\chi^2$ approximation gives more conservative results than the pole method. Since - as we will see in the following sections - using the {\tt pole} method leads to some over-coverage, this is clearly undesirable.

%

\section{\label{sec:cov} Tests of Coverage}

 From a frequentist point of view, an algorithm is said to have the correct {\em coverage\/} if, 
given a confidence level $1-\alpha$ and a large number of repeated
identical experiments, it provides correct answers in a fraction $1-\alpha$ of the cases, independent of the value of $s$. 
To test the coverage of the algorithm proposed in this paper, we
perform the construction described in the previous sections for a large number of simulated experiments, where we
predefine the {\em true\/} signal and background and then determine
$n_o$ by random sampling from a Poisson distribution. 
We then calculate how often the obtained
confidence interval does not contain the predefined $s$. We define the
{\em coverage ratio\/}:
\begin{equation}
R =\frac{n_{false}}{n_{tot}}.
\label{eq:R}
\end{equation}
Here, $n_{false}$ denotes the number of simulated experiments in which the
result of the algorithm does not contain the predefined $s$, and $n_{tot}$ denotes the
number of simulated experiments performed. If we choose $1- \alpha$ to
be 0.9, perfect coverage would  mean $R = 0.1$, independent of signal expectation assumption.\\
A value or $R$ smaller than $0.1$ means that the method over-covers. 
Expected coverage was studied mostly in the context of 
Bayesian intervals, small number of events or including conditioning
~\cite{Narsky:2000} ~\cite{Mandelkern:2001} ~\cite{Helene:99}
and without taking into account systematic uncertainties. Very recently, a study was presented considering coverage including systematic uncertainties ~\cite{Conrad:2002ur}. 

In the next sections, besides presenting coverage tests done for
higher signal expectations without uncertainties, 
we will focus on the coverage of the methods if systematic
uncertainties are included.

\subsection{Coverage without systematic uncertainties}
\begin{figure*}[t]
\begin{minipage}[t]{0.46\linewidth}
\centering\epsfig{file=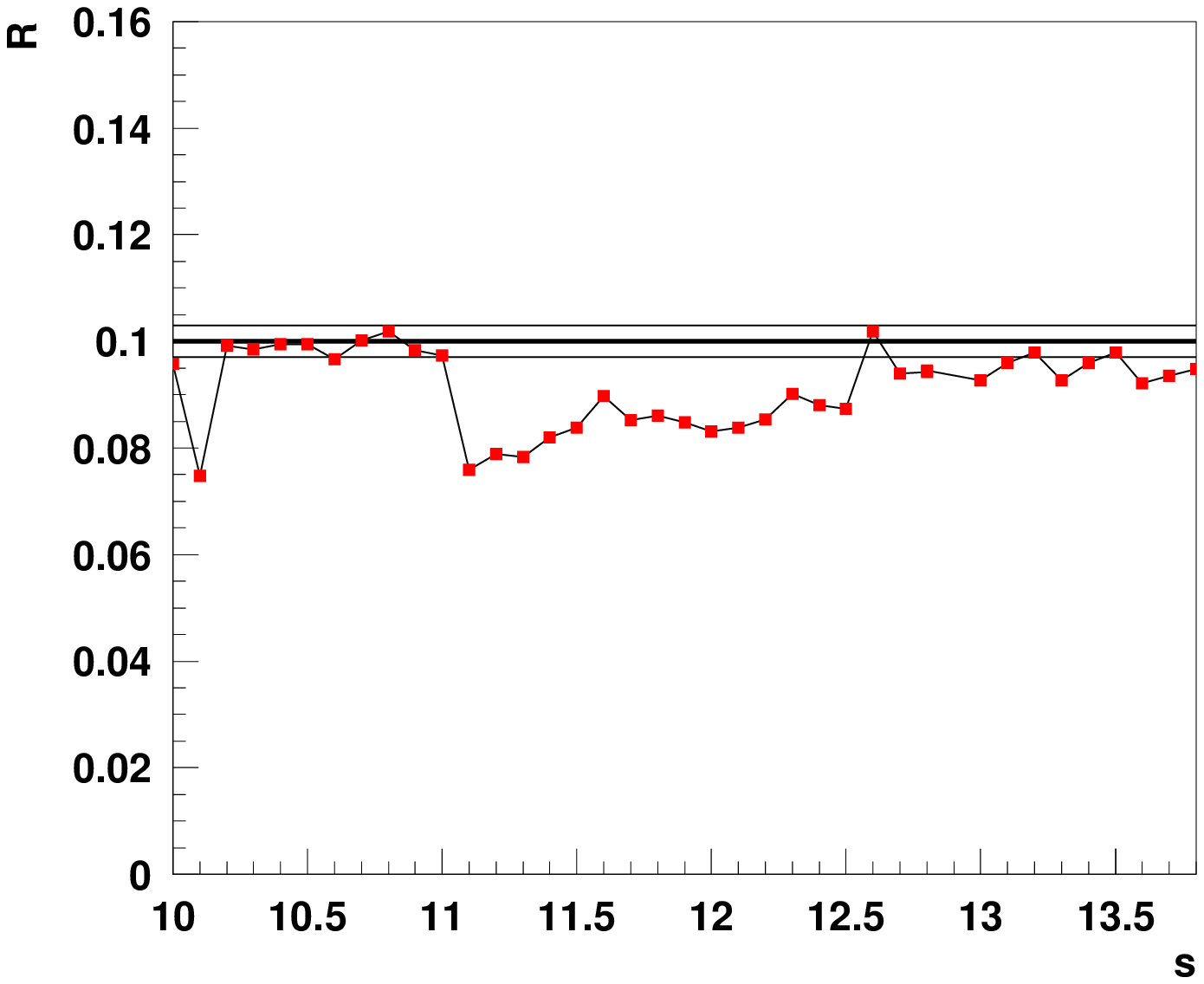,height=\linewidth,width=\linewidth}
\end{minipage}\hfill
\begin{minipage}[t]{0.46\linewidth}
\centering\epsfig{file=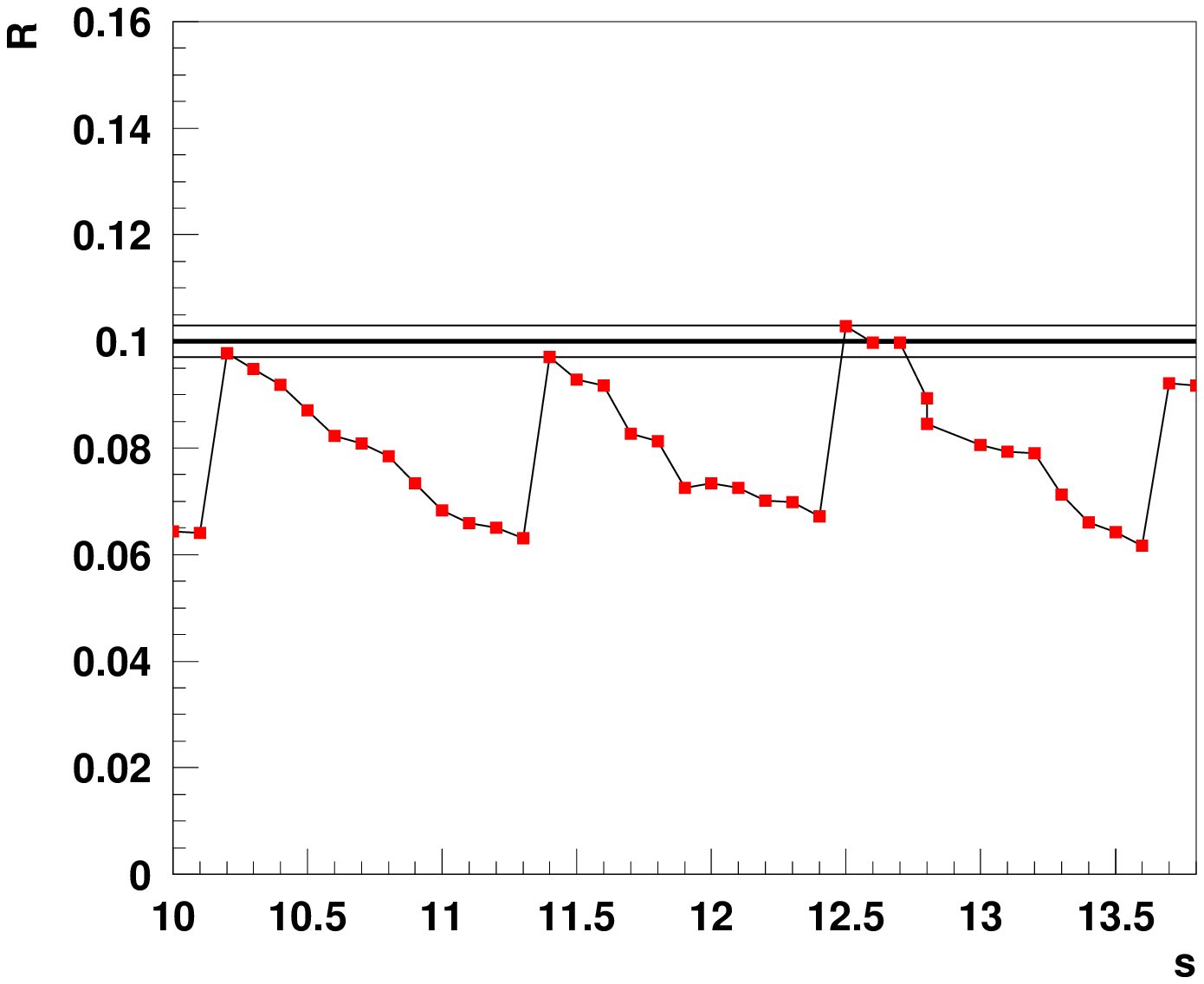,height=\linewidth,width=\linewidth}
\end{minipage}
\caption{Coverage ratio as a function of different signal expectation
assumptions. Left plot: Likelihood ratio ordering. Right plot:
Neyman ordering. 
The thick line gives the line of perfect coverage, the
thinner lines denote the measurement precision of this ratio we can
achieve with 10000 simulated experiments (taken as 1 $\sigma$ of a binomial
distribution). A constant background expectation of b= 10 has been assumed.}
\label{fig:cover1}
\end{figure*}

We show an example of a plot of the coverage ratio (here using steps of 0.1 in signal space) in figure~\ref{fig:cover1} for Neyman and likelihood ratio ordering.
Both methods seem to over-cover for almost all cases (which is expected
because of the discreteness of the Poisson distribution). There is no
signal expectation dependence of the coverage ratio  except for the ``see-saw''
behavior which again reflects the discreteness of the Poisson distribution.

\subsection{Coverage with systematic uncertainties}
\begin{figure*}[t]
\begin{minipage}[t]{0.46\linewidth}
\centering\epsfig{file=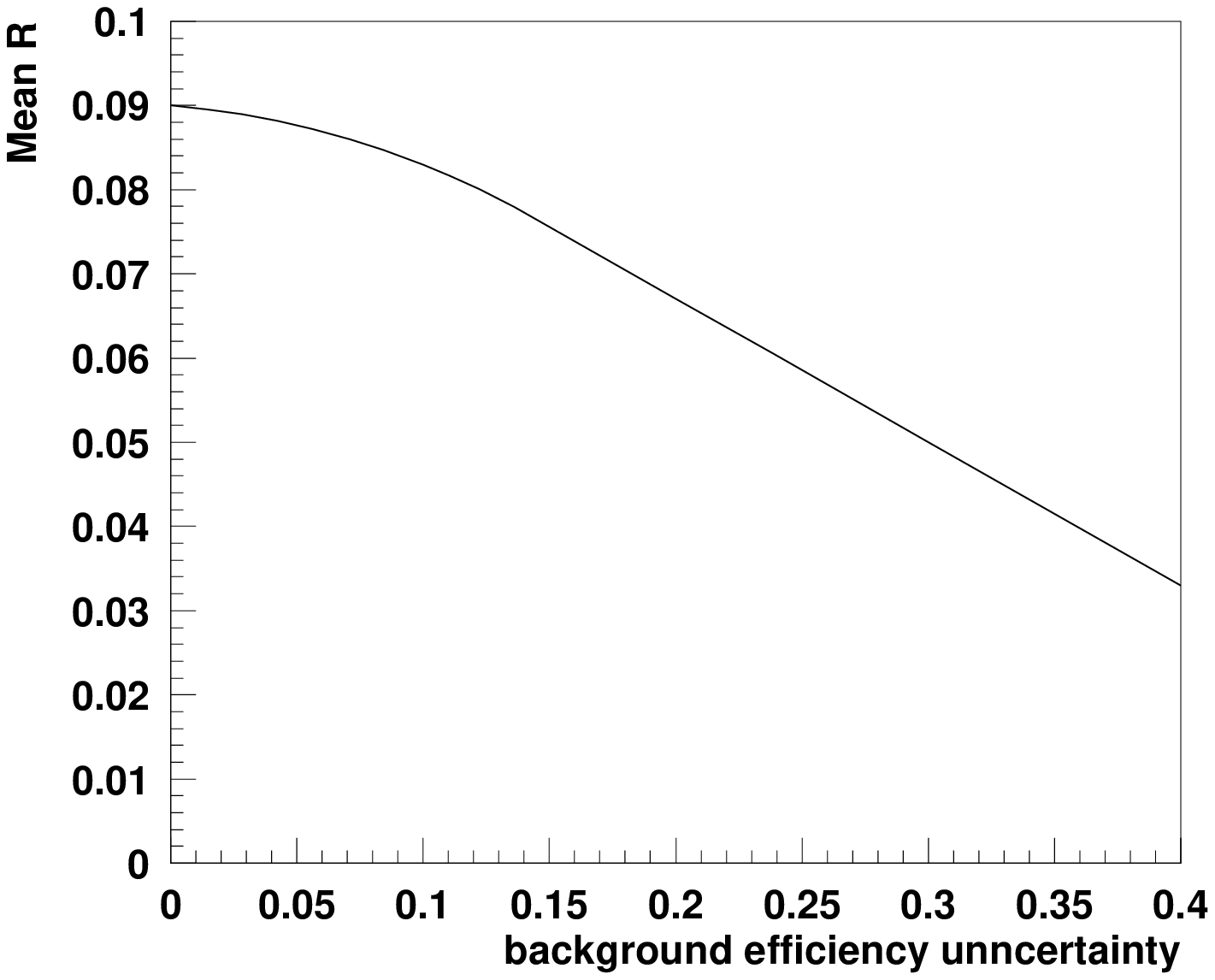,height=\linewidth,width=\linewidth}
\end{minipage}\hfill
\begin{minipage}[t]{0.46\linewidth}
\centering\epsfig{file=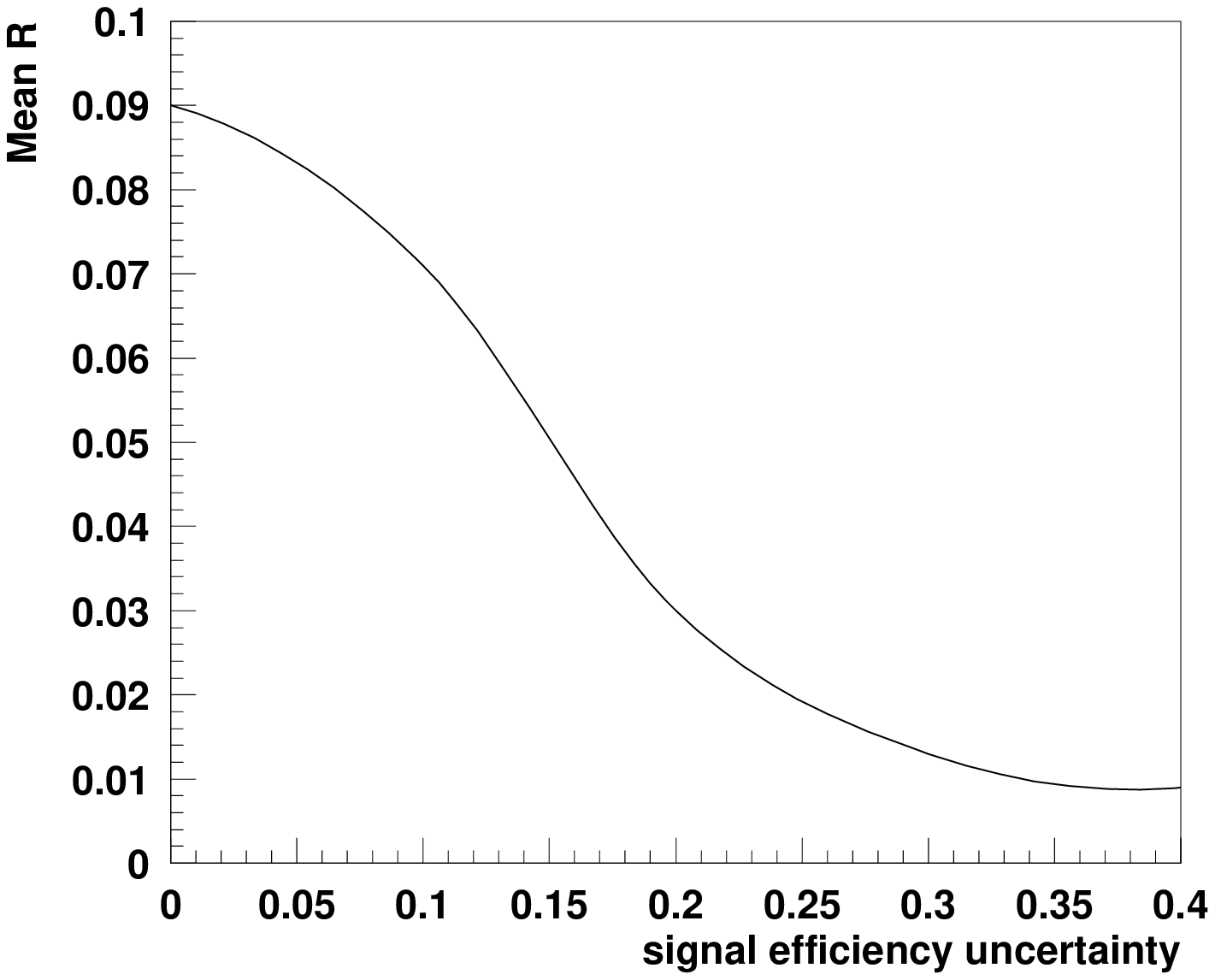,height=\linewidth,width=\linewidth}
\end{minipage}
\caption{Mean coverage ratio as a function of background  
uncertainty (left plot) and signal efficiency uncertainty (right plot).
Here the mean is taken over six signal expectation assumptions
between 12 and 42. The background expectation was taken to be constant
b = 12. All other uncertainties than the one displayed were assumed to be zero.}
\label{fig:cover5}
\end{figure*}

Introducing systematic uncertainties in the calculation of 
confidence intervals and tests of coverage leads to the
question of what is meant by a {\em repeated\/} experiment. If we adhere to
the traditional definition in which an experiment is repeated
with {\em fixed\/} parameters such as efficiency or background
rate, the algorithm presented here will inevitably yield
over-coverage. 
Figure \ref{fig:cover5} shows the mean coverage ratio (mean here
taken over six different signal expectation assumptions) as a function
of different systematic uncertainties. The over-coverage will not only be increasing with increasing uncertainties  but also be dependent on the signal
expectation (figure \ref{fig:cover7}). 

\subsubsection{Bayesian Coverage.}
\begin{figure*}[t]
\begin{minipage}[t]{0.46\linewidth}
\centering\epsfig{file=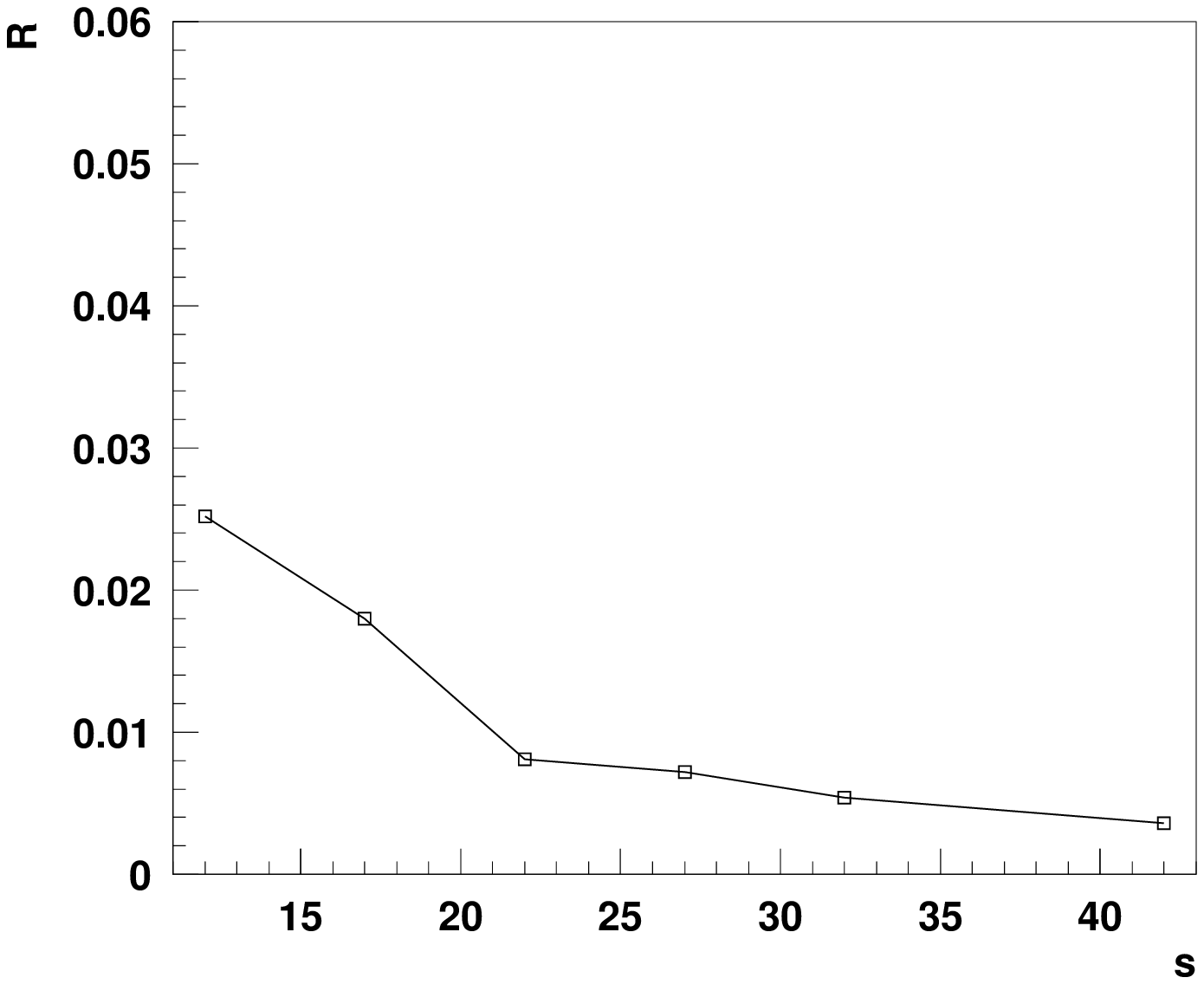,height=\linewidth,width=\linewidth}
\caption{Signal expectation dependence of the coverage ratio. Here for
the case where signal efficiency uncertainty is 30 \%.}
\label{fig:cover7}
\end{minipage}\hfill
\begin{minipage}[t]{0.46\linewidth}
\centering\epsfig{file=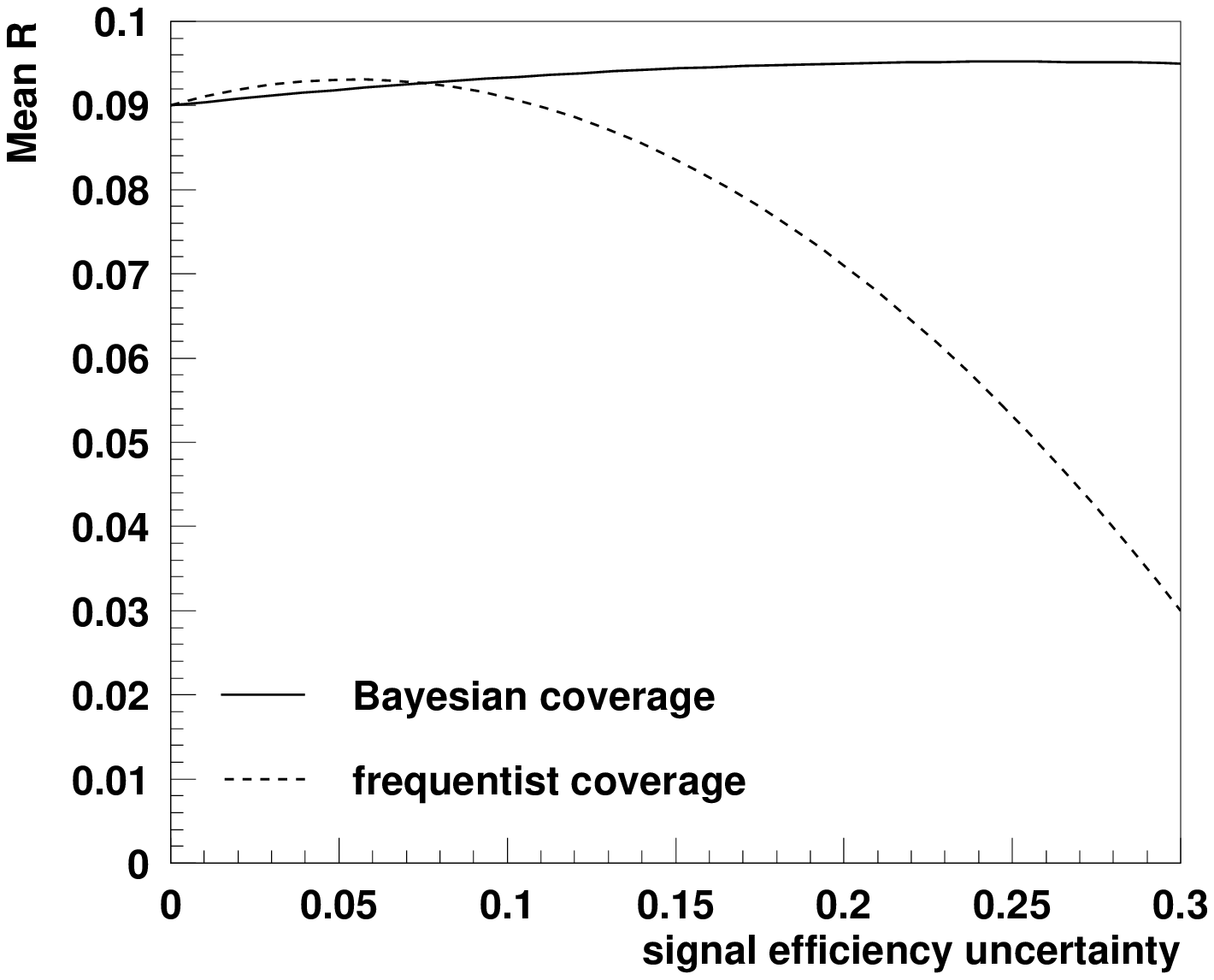,height=\linewidth,width=\linewidth}
\caption{Bayesian mean coverage ratio as a function of the uncertainty
in signal efficiency. The mean is here taken over 29 signal expectation values. For comparison
the frequentist result using the same signal expectation assumptions has been included.}
\label{fig:Bay}
\end{minipage}
\end{figure*}

The  over-coverage described in the previous section is a consequence
of the fact that efficiencies and background are not random variables
(there is a true but unknown fixed efficiency and background rate) but
they are treated as random variables in the construction of the
confidence belt (equations (\ref{eq:b}) and (\ref{eq:epsilon})). \\
Thus, while in the construction we are using a PDF which is a convolution of a Poisson distribution
with a Gaussian distribution, repeated measurements (with parameters fixed)
will produce a Poisson distribution. \\ 
However, one has to keep in mind, that the 
distribution obtained in this way is not the {\em underlying\/} one. To infer from the
measured Poisson distribution the underlying one, the signal efficiency and
the background have to be taken into account. In particular,
if these parameters are uncertain, there will  not be a single 
underlying Poisson distribution, but a set of distributions that
are weighted with the probabilities of the possible different efficiencies and
backgrounds. In a way, we thus give different hypotheses different weights. 
To take this into account we modify the coverage test
described in the previous section. Instead of drawing a measurement 
from Poisson distributions with predefined signal expectation
and background, we draw the signal expectation and background prediction used in each
simulated experiment from Gaussian distributions centered on the
predefined true signal and background, where the width of the Gaussian is the
associated systematic uncertainty. The measurement is then produced by
taking these new values as means for the final Poisson distributions.\\
Since, by using this approach, we give different weights to
different hypotheses, we call this modified coverage test {\em 
Bayesian Coverage\/}. 
In this way, the PDF used in the construction and in the coverage
test are consistent with each other, and the algorithm should, per construction, give
the correct coverage (except for discreteness effects). In particular,
the coverage defined in this way should be independent of the
magnitude of the uncertainties present in the experiment. Figure
\ref{fig:Bay} shows the mean Bayesian coverage for
different  uncertainties in the signal efficiency together with the 
frequentist coverage. The mean is here taken over the 29 points in signal expectation space 
which were tested. As expected the Bayesian coverage ratio is nearly constant.\\
Thus, if we loosen the criteria on the definition of ``repeated experiment'', allowing the ``true'' (unknown) efficiencies vary for each experiment repitition , the method has the desired statistical property of correct coverage.

%

\subsubsection{Remark on the choice of ensemble.}
In the previous subsection we consider an ensemble in which the true value of the nuisance parameter, corresponding to $\epsilon'$ in equation \ref{eq:epsilon}, is varied in each of the members of the ensemble. We find, as expected, that the \texttt{POLE}- method fulfills the requirement of correct coverage with respect to this ensemble.\\
However, it can be argued, that this ensemble does not describe the situation usually encountered in experimental physics. The systematic uncertainty is a {\it measurement} uncertainty, i.e. not the {\it true} value of the nuisance parameter changes in each experiment but the {\it measured} one.\\
Studies for a few cases indicate, that with respect to such an ensemble the \texttt{POLE} method leads to moderate over-coverage \cite{Conrad:2002ur}, \cite{Cousins:2002}.

\section{Limits on High Energy Cosmic Neutrinos.}
\label{sec:app}
In experimental situations where the systematic uncertainties are
 negligible small, limits or 
central confidence intervals can be calculated without evaluating the 
effects of the former. In more general situations, including
 systematic uncertainties in the calculation of the limits is
 essential, since it is a way of incorporating the real sensitivity of
a given experiment to the quantity being measured.\\
 In this section we will consider two real examples taken from
 published results of the AMANDA neutrino telescope, where the program
 POLE was used to include the systematics in the final results.  
\begin{figure*}[t]
\begin{minipage}[t]{0.46\linewidth}
\centering\epsfig{file=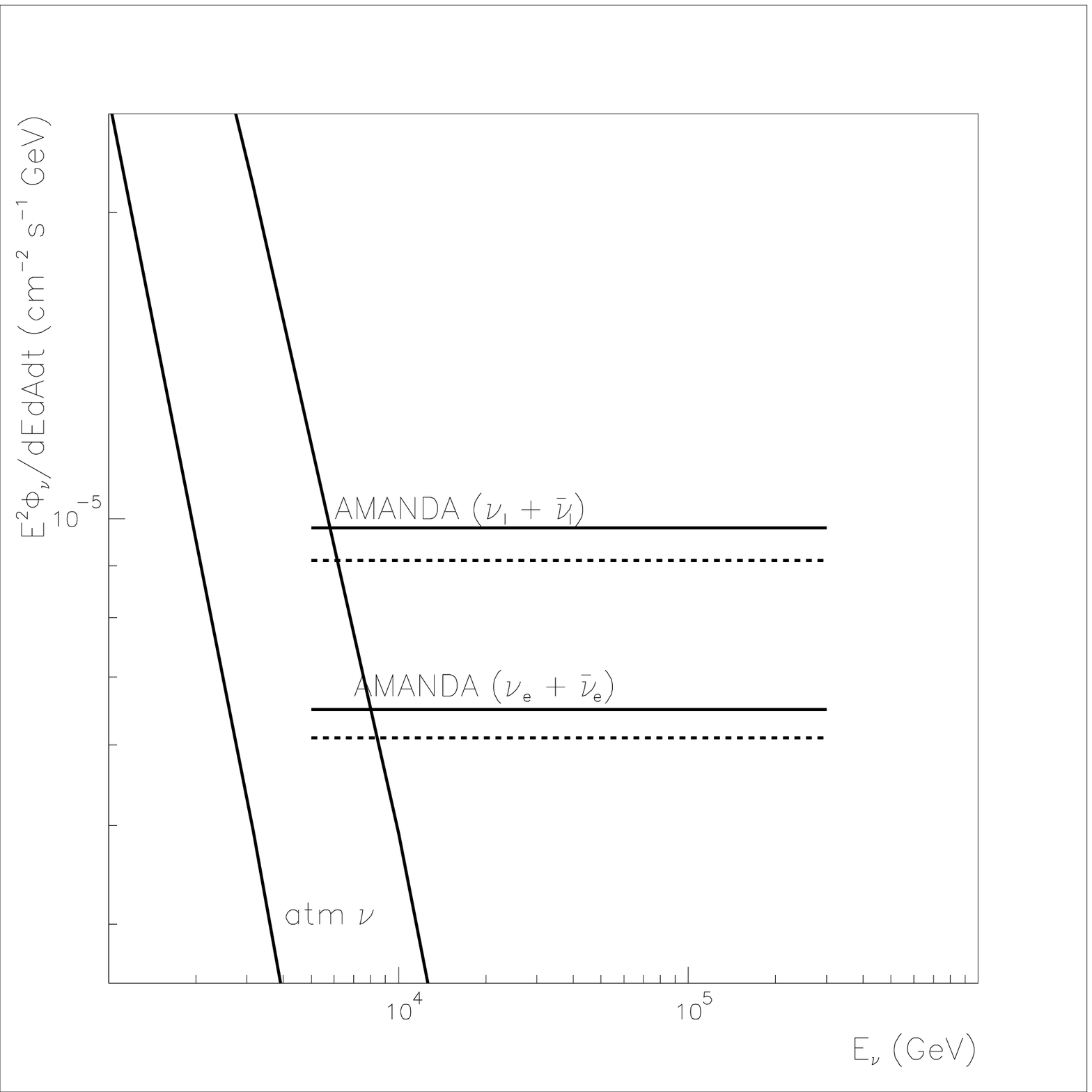,height=\linewidth,width=\linewidth}
\caption{Limit on the flux from cosmic neutrinos of all flavors and electron neutrinos as presented in~\cite{Ahrens:2002wz}. A signal efficiency uncertainty of 25 \% and 30 \% in backround prediction lead to an increase of the upper limit by about 10 \%.}
\label{fig:casc}
\end{minipage}
\end{figure*}
\begin{figure*}[t]
\begin{minipage}[t]{0.46\linewidth}
\centering\epsfig{file=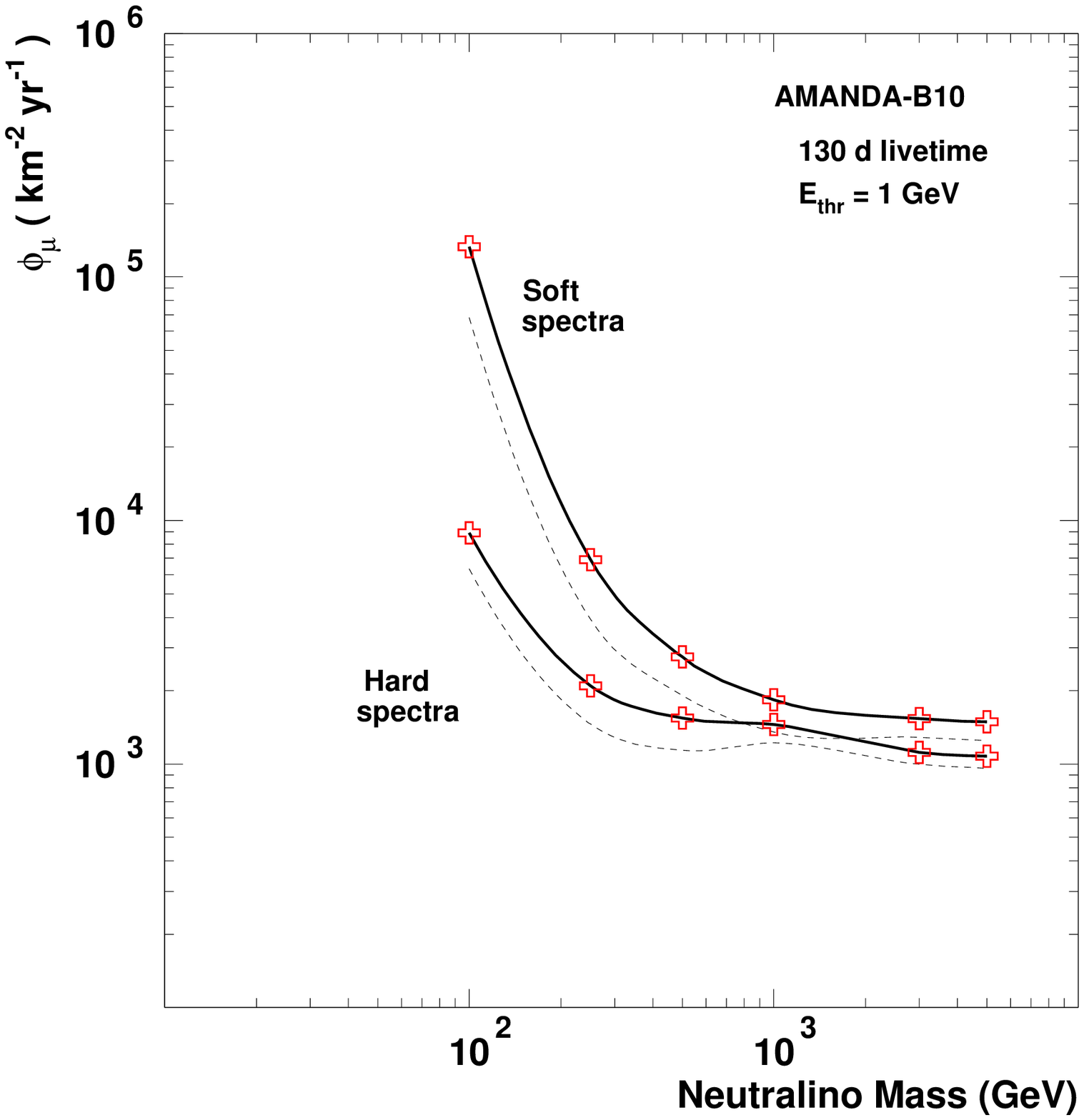,height=\linewidth,width=\linewidth}
\caption{Comparison of the effect of systematics on the limit on the
neutrino-induced muon flux from the center of the Earth from neutralino annihilation. The solid line represents the current limit set by the AMANDA collaboration, the dashed line is the limit without including the systematic uncertainties present in their analysis (figure taken from \cite{Ahrens:2002eb}, where more details of the analysis can be found.)}
\label{fig:WIMP1}
\end{minipage}\hfill
\begin{minipage}[t]{0.46\linewidth}
\centering\epsfig{file=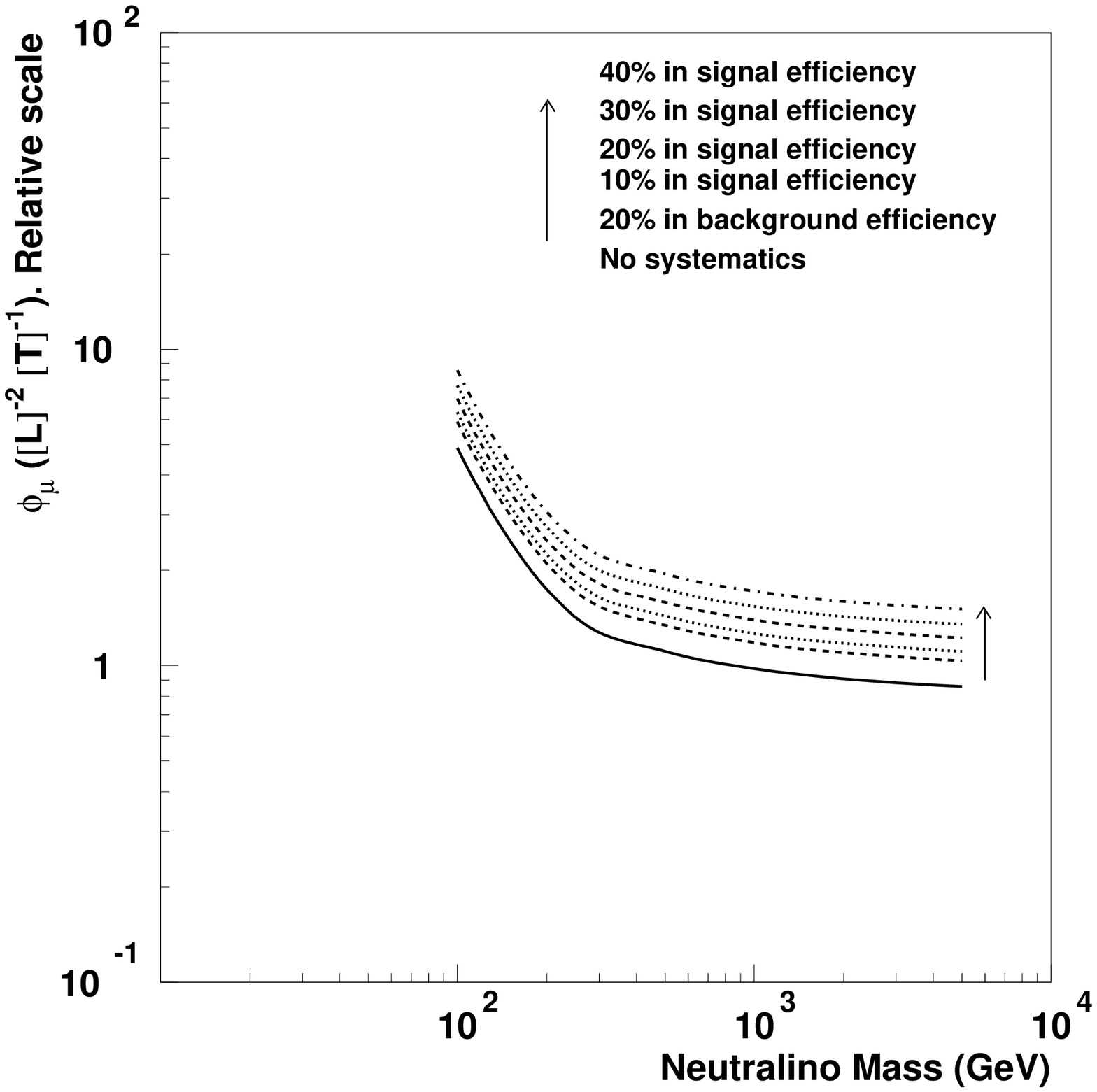,height=\linewidth,width=\linewidth}
\caption{Effect of including uncertainties on a generic WIMP limit. The solid line represents the limit calculation without any experimental systematic uncertainties. The dashed lines represent (from
bottom to top): a 20\% uncertainty in the background expectation only, a
10\%, 20\%, 30\% and 40\% uncertainty on the signal efficiency on top
of the 20\% background uncertainty. Additionally, a 30\% uncertainty in the theoretical atmospheric neutrino background expectation has been assumed in all cases. The absolute scale is arbitrary}
\label{fig:WIMP2}
\end{minipage}
\end{figure*}
 The AMANDA collaboration has published recently 90\% confidence
 limits on the difuse flux of cosmic electron neutrinos and of
 neutrinos all flavours in the energy range between 5 TeV and 300
 TeV \cite{Ahrens:2002wz}. The analysis revealed zero events
 with an expected background from atmospheric neutrinos of 0.01
 events. The systematic uncertainty in the signal efficiency for this 
analysis was $\sim$25\%, determined from Monte Carlo studies. On top
 of that, the current theoretical systematic uncertainty in the atmospheric neutrino
flux prediction in the energy range relevant to the analysis is about
 30\% \cite{Gaisser:2001a}, which has to be taken into account as well. 
 The effect of including the systematic uncertainties is shown in
 figure 12. The effect is to worsen the limits by about 10\%.\par

 Another example worth noticing are the results published by the same 
collaboration on searches for supersymmetric dark matter
   in the form of weakly interacting massive particles,
 WIMPs~\cite{Ahrens:2002eb}. In this analysis the uncertainties in
 signal efficiency range from 10\% to 25\% depending on the assumed 
signal spectrum and, additionally, 
an uncertainty in the background detection efficiency, estimated to be 20\%, 
has to be taken into account in this case.  
A further complication to include these uncertainties in the
calculation of the final limits arises since the efficiencies in
signal and background detection are correlated. Moreover, the
mentioned theoretical uncertainty on the overall normalization of the 
atmospheric neutrino flux has to be added. In figure 13 we show the
limits to the muon flux from the center of the earth as a function of 
WIMP mass. The full lines show the limits for two diferent assumptions
on the signal spectra, and the dashed lines the corresponding limits 
without including systematic uncertainties in the calculations.\par
 
 For the purpose of illustration, we show in figure 14 the effect of 
including systematic uncertainties in the limit calculation for five 
different values (dashed lines). From bottom to top: a 20\% uncertainty in background 
expectation only, and a 10\%, 20\% 30\% and 40\% uncertainty in 
signal efficiency (on top of the mentioned 20\% background
uncertainty). The absolute scale of the plot is arbitrary since we are
just interested in showing the relative effect of the inclusion of 
systematic uncertainties in the limit calculation, with respect to the 
no-systematics case (full line). The figure shows the importance of 
correctly evaluating systematics and including them in the final
result, since the effects can be important.

\section{Conclusions}
\label{sec:con}
In this note we present a Monte Carlo algorithm for introducing systematic 
uncertainties in the evaluation of classical confidence intervals
which allows to include uncertainties in the background prediction, in the background
detection efficiency and in the signal detection efficiency, and
correlations between them, by integrating over the (assumed) PDFs
of these parameters. We apply the method for a Poisson process with background under the assumption of a Gaussian PDF describing the uncertainties. We present results where the construction has been performed using likelihood ratio  ordering with and without conditioning.\\
Generally, the introduction of systematic uncertainties leads to an increase in confidence interval width. However, an interesting result is that likelihood ratio (as well as Neyman) confidence intervals which take into account the systematic uncertainty in the signal efficiency do not become larger with larger uncertainty, in the case that significantly less events have been observed than expected background. 
With respect to an ensemble with strictly identical experiments, introducing systematic uncertainties in the presented manner inevitably leads to over-coverage, increasing with the magnitude of systematic uncertainties. However, we show that with respect to an ensemble where the systematic uncertainties are taken into account in the coverage test by varying their {\it true} assumed values in each pseudoexperiment, the method presented here provides over-coverage only on the level already present due to the discreteness of the Poisson distribution.Both ensembles are ideal ensembles and have to be seen as approximations to the ensemble encountered in experimental physics.
In summary, the algorithm presented here provides a practical and flexible way
to quantitatively take into account systematic uncertainties present
in experimental situations in the calculation of confidence intervals.\\

\begin{acknowledgments}
We are thankful to Robert Cousins for valuable comments on the manuscript. We thank John Conway for making information available in the early stages of this
work. Alexander Biron is acknowledged for careful reading of the manuscript in an earlier version.
\end{acknowledgments}



\bibliographystyle{natbib}

\end{document}